\documentclass[aps,pra,reprint,superscriptaddress]{revtex4-2}
\usepackage{amsmath,amssymb}
\usepackage{graphicx}
\usepackage{siunitx}
\usepackage{hyperref}
\usepackage{orcidlink}

\hypersetup{colorlinks=true,linkcolor=blue,  citecolor=blue,urlcolor=blue}

\begin{document}

\date{\today}

\title{Thermodynamic and Radiative Properties of Euler-Heisenberg AdS Black Holes Surrounded by Quintessence and Dark Matter with a Cloud of Strings}

\newcommand{\RGU}{}
\author{Faizuddin Ahmed\orcidlink{0000-0003-2196-9622}}
\email{faizuddinahmed15@gmail.com}
\affiliation{Department of Physics, The Assam Royal Global University, Guwahati-781035, Assam, India}

\author{Edilberto~O~Silva\orcidlink{0000-0002-0297-5747}}
\email{edilberto.silva@ufma.br}
\affiliation{Programa de P\'{o}s-Gradua\c{c}\~{a}o em F\'{i}sica \& Coordena\c c\~ao do Curso de F\'{\i}sica -- Bacharelado, Universidade Federal do Maranh\~{a}o, 65080-805, S\~{a}o Lu\'{i}s, Maranh\~{a}o, Brazil}

\date{\today}

\begin{abstract}
We investigate the thermodynamics, criticality, and selected radiative and optical properties of an Euler--Heisenberg AdS black hole surrounded by quintessence, perfect fluid dark matter, and a cloud of strings. Within the extended phase-space formalism, we derive the thermodynamic quantities, verify the modified first law and Smarr relation, and analyze the corresponding equation of state and critical behavior. We show that the Euler--Heisenberg coupling and the surrounding matter fields substantially modify the temperature profile, the stability structure, and the location of the critical point. We also examine the sparsity of Hawking radiation, together with the photon sphere, black hole shadow, and the associated geometric-optics emission rate.
\end{abstract}

\maketitle

\section{Introduction}\label{sec:1}

Black hole thermodynamics continues to provide one of the most fruitful bridges between gravitation, quantum theory, and statistical physics. Since the discovery that black holes possess temperature and entropy, they have been understood not merely as classical geometrical objects, but as thermodynamic systems with a rich microscopic structure \cite{Bekenstein1973,BardeenCarterHawking1973,Hawking1975}. In anti-de Sitter (AdS) backgrounds, this picture becomes even more compelling because black holes can exhibit phase transitions analogous to those of ordinary thermodynamic substances. In the extended phase-space approach, where the cosmological constant is identified with the thermodynamic pressure, the analogy with standard fluid systems becomes particularly transparent, leading to the framework now widely known as black-hole chemistry \cite{HawkingPage1983,KastorEtAl2009,KubiznakMann2012,GunasekaranEtAl2012,KubiznakMann2015,KubiznakMannTeo2017}.

A second important direction in the modern study of black holes concerns nonlinear electrodynamics (NLED). Among the best-known effective models is Euler--Heisenberg electrodynamics, which incorporates nonlinear corrections to Maxwell theory and captures quantum effects associated with strong electromagnetic fields \cite{HeisenbergEuler1936,NashedNojiri2021,GuerreroRubiera2020}. Such corrections can substantially modify the near-horizon geometry and influence the thermodynamic observables, especially in the small-black-hole regime where higher-order electromagnetic terms become increasingly relevant. Recent studies of black holes with nonlinear-electrodynamics corrections show that these terms can also leave clear signatures in phase transitions, greybody factors, and optical observables \cite{HamilLutfuoglu2024,AlBadawiAhmed2025,AhmedBlackStrings2025,AhmedPDU2025,AhmedBTZ2025,BretonLopez2021,MagosBreton2020EHAdS,SalazarEtAl1987,HamilLutfuoglu2024,AlBadawiAhmed2025}.

Besides nonlinear electromagnetic effects, it is also of considerable interest to investigate black holes embedded in physically motivated matter environments. Quintessence provides a phenomenological description of dark energy through a fluid with a negative equation-of-state parameter \cite{Kiselev2003}, while perfect fluid dark matter (PFDM) has been used to model the gravitational influence of dark matter halos \cite{LiYang2012}. In addition, a cloud of strings represents another relevant matter source capable of changing the spacetime structure through a deficit-type contribution \cite{Letelier1979}. Each of these external sectors leaves a characteristic imprint on the black hole geometry and, consequently, on its thermodynamic behavior. Closely related studies indicate that perfect-fluid dark matter, quintessence, and string-cloud backgrounds can substantially modify stability windows, criticality, photon orbits, and shadow sizes in a variety of black-hole spacetimes \cite{BezerraEtAl2019,AbbasAli2023,SoodEtAl2024,HeydariFardEtAl2023,HouXuWang2018,HaroonEtAl2019,ShahzadEtAl2025,Rehman2025AccretionEH,Rehman2023AccretionEH,AbbasRehman2023AccretionEHAdS,AcanYildiz2024EH_PFDM,Chaudhary2025EH_PFDM,SucuSakalli2026EHAdS}.

Motivated by these considerations, in this paper, we study an Euler--Heisenberg AdS black hole surrounded simultaneously by quintessence, PFDM, and a cloud of strings. Our aim is to clarify how these distinct contributions modify the thermal properties, phase structure, and Hawking-emission characteristics of the spacetime. In particular, we derive the thermodynamic quantities in the extended phase-space formalism, examine the corresponding criticality conditions, and analyze the sparsity of Hawking radiation. Our discussion is also motivated by recent theoretical and observational interest in black-hole shadows and photon rings, especially after the Event Horizon Telescope results for M87* and Sgr A* \cite{Akiyama2019L1,Akiyama2019L4,Akiyama2022L12,PerlickTsupko2022}.

The paper is organized as follows. In Sec.~\ref{sec:2}, we introduce the black hole geometry and the matter content of the model. In Sec.~\ref{sec:3}, we derive the thermodynamic quantities and verify the modified first law and Smarr relation. In Sec.~\ref{sec:4}, we analyze the equation of state and the corresponding criticality conditions. In Sec.~\ref{sec:5}, we discuss the sparsity of Hawking radiation. In Sec.~\ref{sec:6}, we study the photon sphere and the black hole shadow. In Sec.~\ref{sec:7}, we briefly discuss the energy-emission rate in the geometric-optics regime. Finally, in Sec.~\ref{sec:8}, we summarize the main results and comment on possible extensions of the present work.

Throughout the paper, we work in geometrized units $G=c=1$.

\section{Black Hole Geometry with Quintessence, PFDM, and a Cloud of Strings}\label{sec:2}

In this section, we briefly review the gravitational system under consideration and identify the physical role of each matter contribution entering the metric function. Our goal is to work with a black hole configuration that simultaneously incorporates the effects of nonlinear electrodynamics, quintessence, perfect-fluid dark matter, and a cloud of strings. This setup allows us to examine, within a unified framework, how short-range electromagnetic corrections and long-range environmental effects compete to shape the geometry.

Our starting point is Einstein gravity coupled to the Euler--Heisenberg nonlinear electrodynamics, together with additional matter sectors that describe quintessence, PFDM, and a cloud of strings. The resulting spacetime is static, spherically symmetric, and asymptotically AdS. Since the subsequent thermodynamic analysis depends directly on the lapse function, it is useful to summarize the action and the corresponding energy--momentum components in a compact but physically transparent form.

The action corresponding to Einstein gravity coupled to Euler--Heisenberg electrodynamics reads \cite{HeisenbergEuler1936,SalazarEtAl1987,HamilLutfuoglu2024,AlBadawiAhmed2025}
\begin{equation}
    \mathcal{S}=\frac{1}{4\pi}\int d^4x \sqrt{-g}\,\left(\frac{R}{4}-\mathcal{L}(F)\right)+\mathcal{S}_{\rm DM}+\mathcal{S}_{\rm QF}+\mathcal{S}_{\rm CS},
    \label{action-1}
\end{equation}
where $R$ represents the Ricci scalar curvature, $\mathcal{S}_{\rm DM}$ and $\mathcal{S}_{\rm QF}$ are the actions for dark matter and quintessence, respectively, while $\mathcal{S}_{\rm CS}$ is the action associated with the cloud of strings. The Euler--Heisenberg NLED Lagrangian density is
\begin{equation}
    \mathcal{L}(F)=-F+\frac{a}{2}F^2,
    \label{action-2}
\end{equation}
with \(F=\frac{1}{4}F^{\mu\nu}F_{\mu\nu}\). Here, $a$ is the Euler--Heisenberg parameter that regulates the strength of the nonlinear electrodynamic contribution.

To include string-like objects, we consider the Nambu--Goto action \cite{Letelier1979}
\begin{equation}
    \mathcal{S}_{\rm CS}=\int \sqrt{-\gamma}\,\mathcal{M}\,d\lambda^0\,d\lambda^1
    =\int \mathcal{M}\sqrt{-\frac{1}{2}\Sigma^{\mu \nu}\Sigma_{\mu\nu}}\,d\lambda^0\,d\lambda^1,
    \label{action-4}
\end{equation}
where $\mathcal{M}$ is the dimensionless constant characterizing the string, and $(\lambda^0,\lambda^1)$ are the timelike and spacelike parameters on the world sheet \cite{Goto1971}. The quantity $\gamma$ is the determinant of the induced metric on the string world sheet,
\[
\gamma=g_{\mu\nu}\frac{\partial x^\mu}{\partial \lambda^a}\frac{\partial x^\nu}{\partial \lambda^b},
\]
and
\[
\Sigma_{\mu\nu}=\epsilon^{ab}\frac{\partial x^\mu}{\partial \lambda^a}\frac{\partial x^\nu}{\partial \lambda^b}
\]
is the associated bivector, with $\epsilon^{01}=-\epsilon^{10}=1$.

The corresponding energy--momentum tensor is
\begin{equation}
   T_{\mu\nu}^{\rm CS}
   =2\frac{\partial}{\partial g_{\mu \nu}}
   \left(\mathcal{M}\sqrt{-\frac{1}{2}\Sigma^{\mu \nu}\Sigma_{\mu\nu}}\right)
   =\frac{\rho^{\rm CS}\,\Sigma_{\alpha\nu}\,\Sigma_{\mu}^\alpha}{\sqrt{-\gamma}},
   \label{action-5}
\end{equation}
where $\rho^{\rm CS}$ is the proper density of the cloud of strings. The nonvanishing components are
\begin{equation}
    T^{t\,(\rm CS)}_{t}=\rho^{\rm CS}=\frac{\alpha}{r^2}=T^{r\,(\rm CS)}_{r}, \qquad
    T^{\theta\,(\rm CS)}_{\theta}=T^{\phi\,(\rm CS)}_{\phi}=0,
    \label{action-6}
\end{equation}
where $\alpha$ is the string cloud parameter.

Quintessence is modeled following Kiselev \cite{Kiselev2003} as an anisotropic fluid with equation-of-state parameter $\omega_q\in(-1,-1/3)$ and energy--momentum components
\begin{align}
T^{t\,(\rm QF)}_{t}=T^{r\,(\rm QF)}_{r}
=-\frac{3N\omega_q}{2\,r^{3(1+\omega_q)}},\\
T^{\theta\,(\rm QF)}_{\theta}
=-\frac{3\omega_q+1}{2}\,T^{t\,(\rm QF)}_{t},
\label{action-7}
\end{align}
with $N>0$ denoting the quintessence intensity.

Considering dark matter as a perfect fluid, its energy-momentum tensor can be written as
\begin{equation}
T_{\mu\,(\rm DM)}^{\nu}=\mbox{diag}\left(-\rho,p,p,p\right),
\label{action-8}
\end{equation}
together with the angular trace condition
\[
T^{\theta(\rm DM)}_{\theta}=(1-\delta)\,T^{t(\rm DM)}_{t},
\]
and the equation of state
\begin{equation}
p=(\delta-1)\rho,
\label{action-9}
\end{equation}
where $\delta=3/2$ is the standard choice that yields the logarithmic tail associated with flat galactic rotation curves \cite{LiYang2012}.

The modified field equations can be written schematically as \cite{Ruffini2013EEH}
\begin{equation}
R_{\mu \nu }-\frac{1}{2}g_{\mu \nu }R
=8\pi\left(\bar{T}_{\mu \nu}+4\pi T^{(\rm DM)}_{\mu \nu}-T^{(\rm QF)}_{\mu \nu }\right),
\label{EFEs}
\end{equation}
where $\bar{T}_{\mu\nu}$ denotes the effective contribution coming from the Euler--Heisenberg electromagnetic sector.

The line element of a static and spherically symmetric spacetime is described by
\begin{equation}
    ds^2=-f(r)\,dt^2+\frac{dr^2}{f(r)}+r^2(d\theta^2+\sin^2\theta\,d\varphi^2),
    \label{metric}
\end{equation}
where the lapse function is given by
\begin{equation}
     f(r)=1-\alpha-\frac{2M}{r}+\frac{Q^2}{r^2}-\frac{aQ^4}{20r^6}
     -\frac{N}{r^{3\omega_q+1}}+\frac{\lambda}{r}\ln\!\frac{r}{|\lambda|}
     +\frac{r^2}{\ell^2}.
\label{function}
\end{equation}
Here, $M$ represents the mass parameter, $Q$ is the electric charge, $\alpha$ is the cloud-of-strings parameter \cite{Letelier1979}, $(N,\omega_q)$ characterize the quintessence background \cite{Kiselev2003}, $\lambda$ is the PFDM parameter \cite{LiYang2012}, and $\ell$ is the AdS radius.

\begin{figure}[tbhp]
    \centering
    \includegraphics[width=\linewidth]{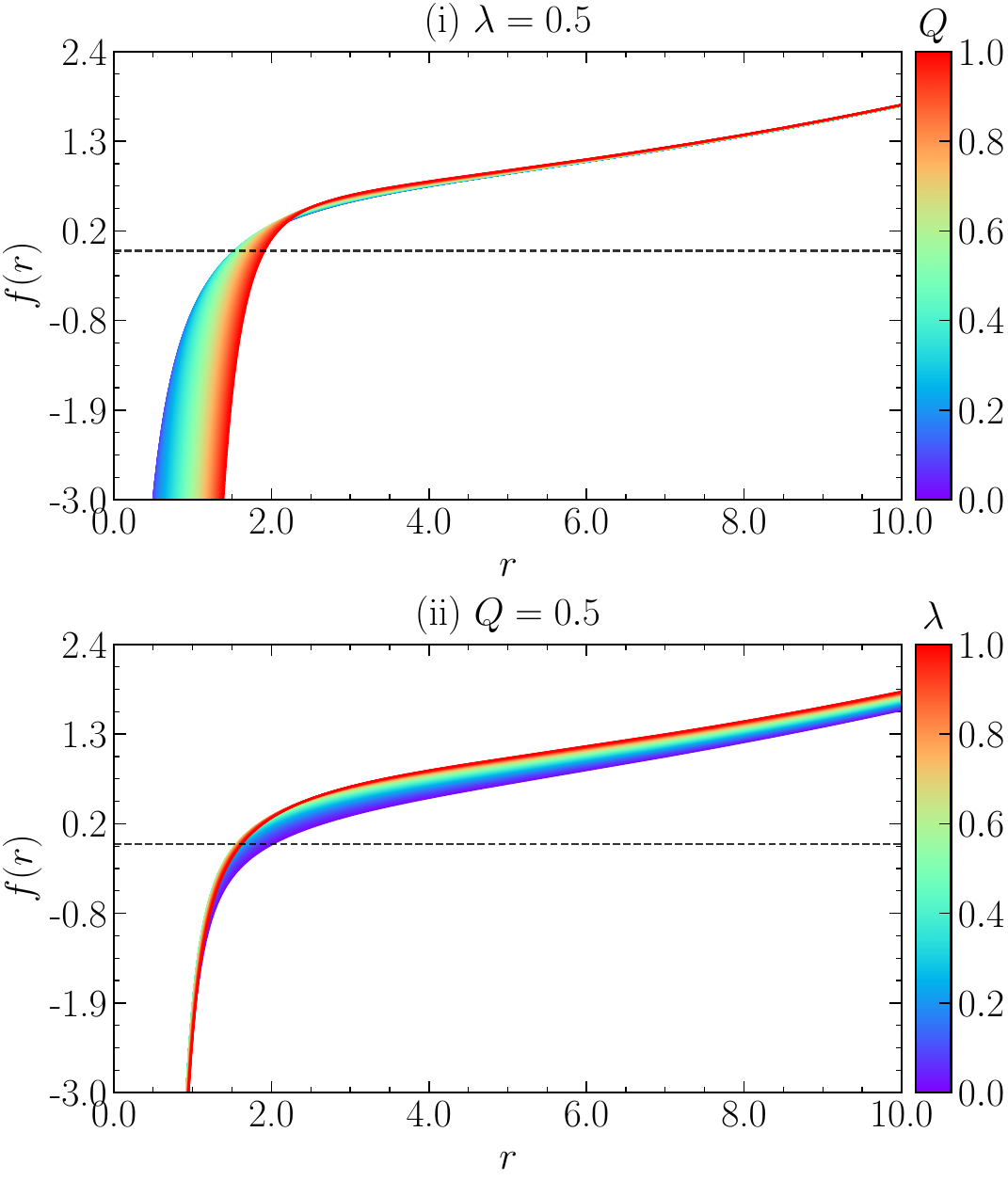}
    \caption{Two-panel behavior of the metric function $f(r)$ as a function of the radial coordinate $r$. In the upper panel, the PFDM parameter is fixed at $\lambda=0.5$, and the electric charge $Q$ is varied. In the lower panel, the charge is fixed at $Q=0.5$ and $\lambda$ is varied. The remaining parameters are $M=1$, $\alpha=0.1$, $P=\tfrac{0.03}{8\pi}$, $N=0.01$, $\omega_q=-2/3$, and $a=0.50\times10^3$. The intersections of the curves with the horizontal line $f(r)=0$ determine the horizon radii.}
    \label{fig:metric}
\end{figure}
Figure~\ref{fig:metric} displays the behavior of the metric function for two representative families of parameters. Since the event horizon radius is determined by the condition $f(r_h)=0$, the zeros of the curves provide a direct visualization of the horizon structure of the spacetime. In the upper panel, varying the charge $Q$ changes both the depth and the location of the minimum of $f(r)$, thereby shifting the horizon position. In the lower panel, varying the PFDM parameter $\lambda$ also deforms the metric function and shifts the radial position where the horizon condition is satisfied. Therefore, the figure offers a useful geometric interpretation of the thermodynamic analysis, since the horizon radius entering the temperature, heat capacity, and sparsity parameter is obtained precisely from the roots of $f(r)$.

The structure of the lapse function makes the origin of each contribution explicit. The mass term and the electric charge term reproduce the familiar Reissner--Nordstr\"om-type behavior, while the Euler--Heisenberg correction introduces a higher-order inverse-power contribution that becomes important near the horizon. The quintessence term contributes via a power-law decay controlled by the state parameter; the PFDM sector introduces a logarithmic modification associated with halo-like matter effects; and the string cloud parameter shifts the constant part of the metric function. The AdS contribution, as usual, dominates at large distances and is responsible for the pressure term in the extended thermodynamic description.

\section{Thermodynamics}\label{sec:3}

We now turn to the thermodynamic description of the black hole in the extended phase-space framework \cite{KastorEtAl2009,KubiznakMann2012,GunasekaranEtAl2012,KubiznakMannTeo2017}. In this approach, the cosmological constant is promoted to a thermodynamic variable and identified with the pressure, allowing a closer analogy between black hole mechanics and ordinary thermodynamics. This viewpoint is especially useful for studying phase transitions and critical phenomena in AdS spacetimes.

Once the event horizon radius is specified through the horizon condition, all relevant thermodynamic quantities can be expressed in terms of $r_h$ or, equivalently, in terms of the entropy. The presence of the Euler--Heisenberg parameter, quintessence, PFDM, and the cloud of strings modifies the standard AdS black hole expressions in a nontrivial way, so it is important to display not only the formal formulas but also their physical interpretation.

\begin{figure}[tbhp]
    \centering
    \includegraphics[width=\linewidth]{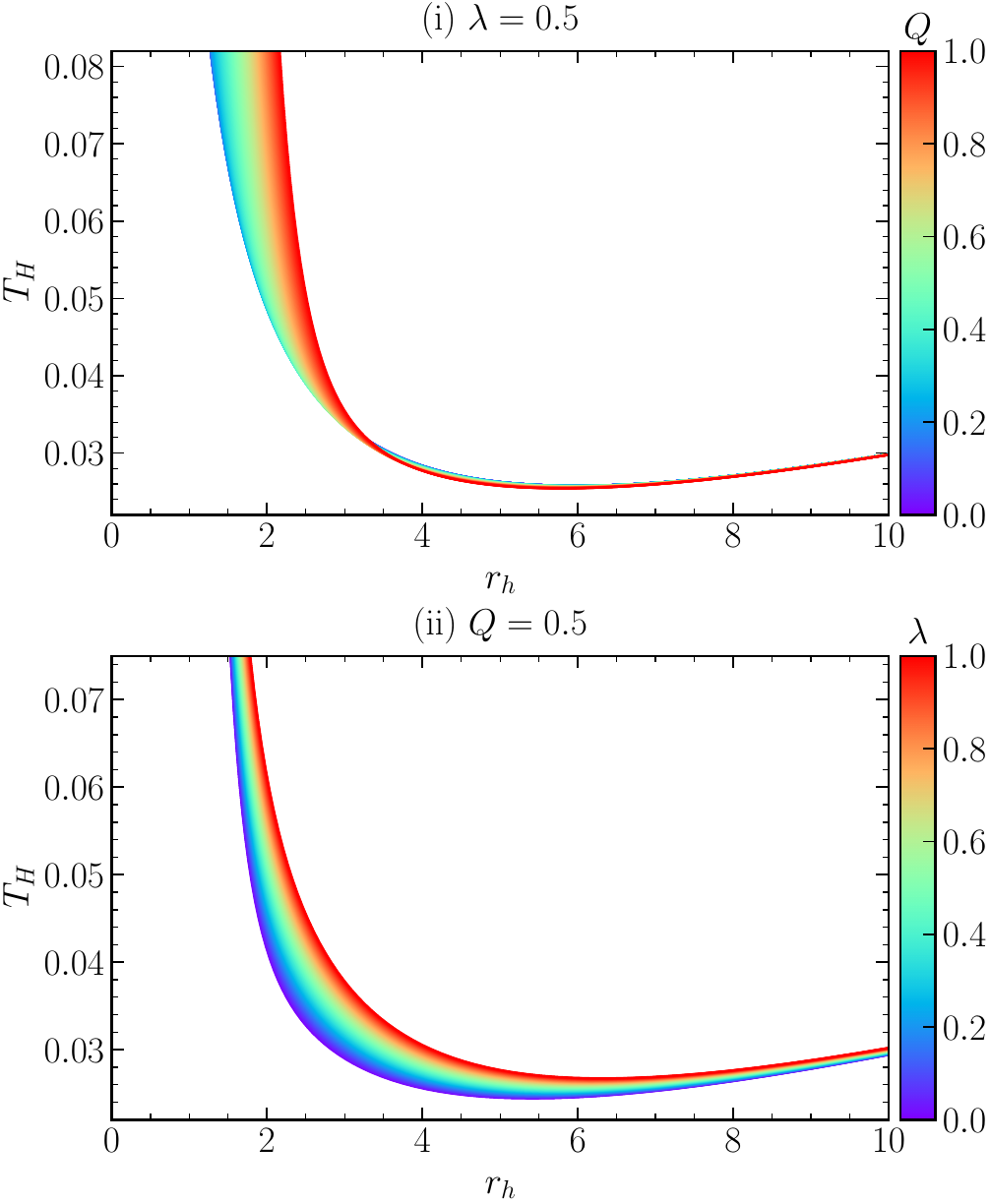}
    \caption{Two-panel behavior of the Hawking temperature $T_H$ as a function of the horizon radius $r_h$. In the upper panel, the PFDM parameter is fixed at $\lambda=0.5$, and the charge $Q$ is varied; in the lower panel, the charge is fixed at $Q=0.5$ and $\lambda$ is varied. The remaining parameters are $\alpha=0.1$, $P=\tfrac{0.03}{8\pi}$, $N=0.01$, $\omega_q=-2/3$, and $a=0.50\times 10^3$.}
    \label{fig:temperature}
\end{figure}

In extended phase space, the thermodynamic pressure $P$ is related to the AdS radius through
\begin{equation}
    \frac{1}{\ell^2}=-\frac{\Lambda}{3}=\frac{8\pi P}{3}.
    \label{aa}
\end{equation}

For later convenience, we introduce the shorthand radial functions
\begin{align}
\mathcal{H}(r)&\equiv 1-\alpha+\frac{Q^2}{r^2}-\frac{aQ^4}{20r^6}
-\frac{N}{r^{3\omega_q+1}}+\frac{\lambda}{r}\ln\!\frac{r}{|\lambda|}\notag\\&
+\frac{8\pi P}{3}r^2,\label{aa0a}\\
\Delta(r)&\equiv 1-\alpha-\frac{Q^2}{r^2}+\frac{aQ^4}{4r^6}
+\frac{3\omega_q N}{r^{3\omega_q+1}}+\frac{\lambda}{r}
+8\pi P r^2,\label{aa0b}\\
\Xi(r)&\equiv 1-\alpha-\frac{Q^2}{r^2}+\frac{aQ^4}{4r^6}
-\frac{(3\omega_q+1)N}{r^{3\omega_q+1}}\notag\\&
+\frac{\lambda}{r}\left(1+\ln\!\frac{r}{|\lambda|}\right)
+8\pi P r^2,\label{aa0c}\\
\Upsilon(r)&\equiv -(1-\alpha)+\frac{3Q^2}{r^2}
-\frac{7aQ^4}{4r^6}
-\frac{3\omega_q(3\omega_q+2)N}{r^{3\omega_q+1}}\notag\\&
-\frac{2\lambda}{r}+8\pi P r^2.\label{aa0d}
\end{align}

The horizon condition $f(r_h)=0$ implies
\begin{align}
M=\frac{r_h}{2}\,\mathcal{H}(r_h).\label{aa1}
\end{align}

The Hawking temperature $T_H=\kappa/(2\pi)$, where $\kappa=f'(r_h)/2$, is given by
\begin{align}
T_H=\frac{1}{4\pi r_h}\,\Delta(r_h).\label{aa2}
\end{align}

The Hawking temperature shows clearly how the different sectors compete with one another. The charge contribution tends to lower the temperature, whereas the pressure term raises it for sufficiently large horizon radius. The Euler--Heisenberg correction contributes with a higher-order inverse-power term and therefore becomes especially relevant for small black holes. The quintessence, PFDM, and string-cloud parameters further deform the temperature profile, potentially altering the location of extrema and, consequently, the thermal stability of the system. Figure~\ref{fig:temperature} makes this interplay explicit: in both panels, the temperature starts from zero at the smallest allowed horizon, rises to a local maximum, and then decreases for larger $r_h$, while variations of $Q$ or $\lambda$ reshape both the height of the peak and the position of the turning point.

The Bekenstein--Hawking entropy is one quarter of the horizon area,
\begin{equation}
    S=\pi r_h^2.
    \label{aa3}
\end{equation}

The Gibbs free energy of the system is
\begin{align}
    G=M-T_HS=\frac{r_h}{4}\,\Big[2\mathcal{H}(r_h)-\Delta(r_h)\Big].
    \label{aa4}
\end{align}

The Gibbs free energy is the most convenient quantity for discussing global thermodynamic preference and possible first-order phase transitions \cite{KubiznakMann2012,KubiznakMann2015}. Its temperature dependence can be used to identify possible first-order phase transitions through cusp- or swallowtail-like structures. In the present model, such behavior is expected to be highly sensitive to the external matter content and the nonlinear electrodynamics parameter, since both enter directly into the mass and temperature functions.

\begin{figure}[tbhp]
\centering
\includegraphics[width=\linewidth]{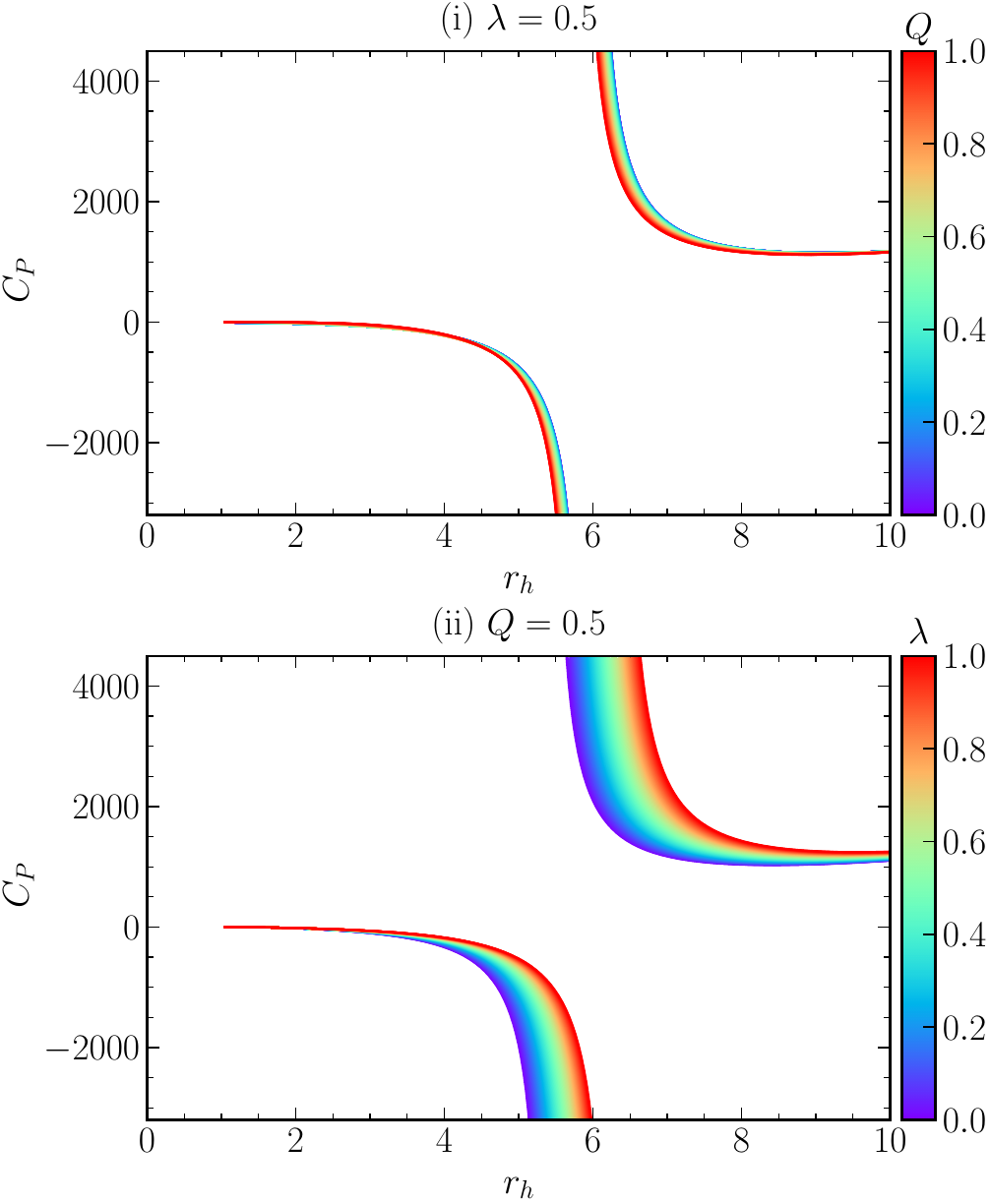}
\caption{Two-panel behavior of the specific heat capacity $C_P$ as a function of the horizon radius $r_h$. In the upper panel, $\lambda=0.5$ is fixed, and the charge $Q$ is varied; in the lower panel, $Q=0.5$ is fixed, and $\lambda$ is varied. The remaining parameters are $\alpha=0.1$, $P=\tfrac{0.03}{8\pi}$, $N=0.01$, $\omega_q=-2/3$, and $a=0.50\times 10^3$.}
\label{fig:heat}
\end{figure}

The heat capacity at constant pressure is given by
\begin{equation}
C_P=\frac{dM}{dT_H}=2\pi r_h^2\,\frac{\Xi(r_h)}{\Upsilon(r_h)}.\label{aa5}
\end{equation}

The heat capacity provides local information about thermal stability. In particular, divergences in $C_P$ mark second-order phase-transition points, separating branches of stable and unstable black hole configurations. Because the denominator contains contributions from the charge, the Euler--Heisenberg parameter, the quintessence sector, PFDM, and the pressure, the stability windows can differ significantly from those of the standard RN-AdS case. Figure~\ref{fig:heat} shows that the locations of the divergences shift appreciably as either $Q$ or $\lambda$ varies, so the size and position of the stable branch are controlled jointly by the gauge sector and the surrounding matter distribution.

In terms of entropy, the mass parameter can be rewritten as
\begin{align}
    M=\frac{1}{2}\sqrt{\frac{S}{\pi}}\,
    \mathcal{H}\!\left(\sqrt{\frac{S}{\pi}}\right).
    \label{aa6}
\end{align}

The modified first law of thermodynamics can then be expressed as
\begin{equation}
dM=T\,dS+V\,dP+\Phi\,dQ+\Psi\,dN+\chi\,d\lambda,
    \label{aa7}
\end{equation}
where the temperature $T$ is
\begin{align}
T=\left(\frac{\partial M}{\partial S}\right)_{P,Q,N,\lambda}
=\frac{1}{4\sqrt{\pi S}}\,
\Delta\!\left(\sqrt{\frac{S}{\pi}}\right)=T_H,
\label{aa8}
\end{align}
the thermodynamic volume $V$ is
\begin{equation}
V=\left(\frac{\partial M}{\partial P}\right)_{S,Q,N,\lambda}
=\frac{4}{3}\sqrt{\frac{S^3}{\pi}}
=\frac{4\pi}{3}r_h^3,
\label{aa9}
\end{equation}
the electric potential $\Phi$ is
\begin{align}
\Phi &=\left(\frac{\partial M}{\partial Q}\right)_{P,S,N,\lambda}\notag\\ &
=\sqrt{\frac{\pi}{S}}\,Q-\frac{aQ^3}{10}\sqrt{\frac{\pi^5}{S^5}}\notag\\&
=\frac{Q}{r_h}-\frac{aQ^3}{10r_h^5},
\label{aa10}
\end{align}

the potential associated with quintessence is
\begin{equation}
    \Psi=\left(\frac{\partial M}{\partial N}\right)_{P,S,Q,\lambda}
    =-\frac{1}{2}\left(\frac{\pi}{S}\right)^{3\omega_q/2},
    \label{aa11}
\end{equation}
and the potential associated with PFDM is
\begin{align}
\chi&=\left(\frac{\partial M}{\partial \lambda}\right)_{P,S,Q,N}
    =\frac{1}{2}\left[
    \ln\!\frac{\sqrt{S/\pi}}{|\lambda|}-1
    \right]\notag\\&
    =\frac{1}{2}\left(
    \ln\!\frac{r_h}{|\lambda|}-1
    \right).
    \label{aa12}
\end{align}

With these quantities, one can verify the modified Smarr relation
\begin{equation}
    M=2TS-2VP+\Phi Q+(3\omega_q+1)\Psi N+\chi\lambda.
    \label{aa13}
\end{equation}

The above expressions confirm that the extended thermodynamic description remains consistent even in the presence of nonlinear electrodynamics and additional matter sources. The modified first law and Smarr relation show that the quintessence intensity and PFDM parameter should be treated as genuine thermodynamic variables, each with its own conjugate potential. This is an important result because it provides a coherent framework for discussing how environmental matter fields affect black hole thermodynamics beyond purely geometric modifications.

\section{Thermodynamic Criticality}\label{sec:4}

Having established the basic thermodynamic quantities, we now investigate the critical behavior of the system, following the standard extended-phase-space strategy widely used in AdS black-hole thermodynamics \cite{KubiznakMann2012,GunasekaranEtAl2012,KubiznakMannTeo2017}. The main objective is to determine whether the black hole undergoes a small/large black hole phase transition analogous to the liquid--gas transition of a Van der Waals fluid. To this end, we rewrite the pressure in terms of the temperature and the horizon radius and then express the equation of state in terms of the specific volume.

For the representative choice $\omega_q=-2/3$, the equation of state takes a form that contains several competing inverse-power contributions. This is particularly useful because each parameter leaves a qualitatively distinct signature: the string cloud modifies the effective $1/v^2$ term, PFDM introduces a $1/v^3$ correction, the electric charge contributes through the $1/v^4$ term, and the Euler--Heisenberg nonlinear correction appears through the higher-order $1/v^8$ contribution. Therefore, the phase structure of the system can be interpreted directly from the hierarchy of terms in the equation of state.

The thermodynamic pressure as a function of temperature for $\omega_q=-2/3$ is
\begin{equation}
    P=\frac{T}{2r_h}
    -\frac{1-\alpha}{8\pi r_h^2}
    +\frac{Q^2}{8\pi r_h^4}
    -\frac{aQ^4}{32\pi r_h^8}
    +\frac{N}{4\pi r_h}
    -\frac{\lambda}{8\pi r_h^3}.
    \label{cc1}
\end{equation}

Transforming to the specific volume $v=2r_h$, we obtain
\begin{equation}
    P=\frac{T}{v}
    -\frac{c_2}{v^2}
    +\frac{c_4}{v^4}
    -\frac{c_5}{v^8}
    +\frac{c_1}{v}
    -\frac{c_3}{v^3},
    \label{cc2}
\end{equation}
where
\begin{align}
    c_1=\frac{N}{2\pi},
    c_2=\frac{1-\alpha}{2\pi},    c_3=\frac{\lambda}{\pi},
    c_4=\frac{2Q^2}{\pi},
    c_5=\frac{8aQ^4}{\pi}.
    \label{cc3}
\end{align}

This form of the equation of state makes the physical content of the model particularly transparent, and it is well suited for comparison with other matter-dressed or nonlinear-electrodynamics AdS black holes studied in the recent literature \cite{BezerraEtAl2019,AbbasAli2023,SoodEtAl2024,AlBadawiAhmed2025}. The coefficient $c_1$ shifts the effective linear term in $1/v$, the coefficient $c_2$ governs the attractive contribution analogous to the usual Van der Waals term, $c_3$ encodes the PFDM correction, $c_4$ is associated with the electric charge, and $c_5$ contains the Euler--Heisenberg nonlinear electrodynamics effect. Since the last term scales as $v^{-8}$, it mainly affects the short-volume regime and is therefore expected to modify the critical point most strongly for small black holes.

The critical point $(v_c,P_c,T_c)$ is obtained from the standard inflection-point conditions,
\[
\left(\frac{\partial P}{\partial v}\right)_T=0,
\qquad
\left(\frac{\partial^2 P}{\partial v^2}\right)_T=0.
\]
The resulting relations are
\begin{align}
&c_2 v_c^6 + 3c_3 v_c^5 - 6c_4 v_c^4 + 28c_5 = 0,\label{cc4}\\
&T_c =-\,c_1+ \frac{2c_2}{v_c}+ \frac{3c_3}{v_c^2}- \frac{4c_4}{v_c^3}+ \frac{8c_5}{v_c^7},\label{cc5}\\
&P_c =\frac{c_2}{v_c^2}+ \frac{2c_3}{v_c^3}- \frac{3c_4}{v_c^4}+ \frac{7c_5}{v_c^8}.\label{cc6}
\end{align}

The critical point is determined by the standard inflection conditions on the isotherms, and the resulting algebraic equation for $v_c$ reflects the combined influence of all matter and interaction parameters. Although a closed analytical expression is not available in the fully general Euler--Heisenberg case, the structure of the equation already shows that the nonlinear-electrodynamics term can significantly shift the critical volume. Once $v_c$ is known, the corresponding critical temperature and pressure follow immediately.

\begin{figure}[bhp]
    \centering
    \includegraphics[width=0.95\linewidth]{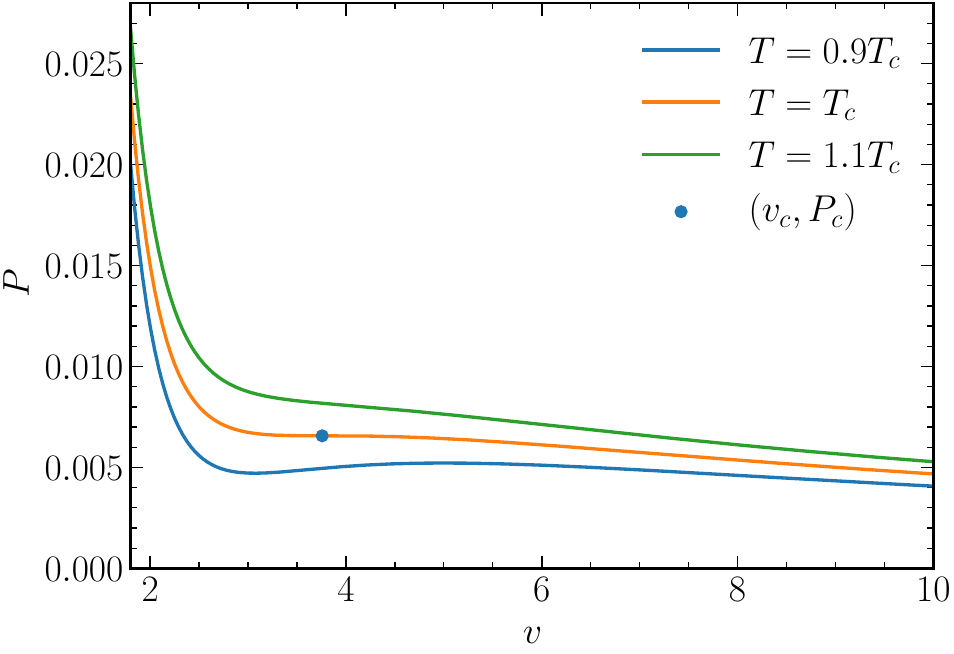}
    \caption{Equation-of-state isotherms $P(v)$ for $T=0.9T_c$, $T=T_c$, and $T=1.1T_c$ with $\alpha=0.1$, $a=0.01$, $Q=1$, $N=0.01$, and $\lambda=0.5$. The critical point $(v_c,P_c)$ is marked by the black dot. The subcritical isotherm displays the characteristic oscillatory Van der Waals behavior, while the critical isotherm develops the expected inflection point.}
    \label{fig:pv}
\end{figure}

Figure~\ref{fig:pv} illustrates this behavior for representative isotherms. Below the critical temperature, the system exhibits the familiar oscillatory structure associated with the coexistence of small and large black hole phases; at $T=T_c$, the inflection point appears; above the critical temperature, the isotherms become monotonic, indicating the disappearance of first-order phase coexistence.

\begin{table*}[tbhp]
\centering
\small
\begin{tabular}{|c|ccccccccc|}
\hline
$N \backslash \lambda$ 
& 0.1 & 0.2 & 0.3 & 0.4 & 0.5 & 0.6 & 0.7 & 0.8 & 0.9 \\
\hline

\multicolumn{10}{c}{$v_c$} \\
\hline
0.01 & 3.948 & 3.697 & 3.465 & 3.249 & 3.050 & 2.867 & 2.699 & 2.544 & 2.402 \\
\hline

\multicolumn{10}{c}{$T_c$} \\
\hline
0.01 & 0.04935 & 0.06105 & 0.07773 & 0.1018 & 0.1368 & 0.1879 & 0.2622 & 0.3699 & 0.5247 \\
0.02 & 0.04776 & 0.05946 & 0.07614 & 0.1002 & 0.1352 & 0.1863 & 0.2606 & 0.3683 & 0.5231 \\
0.03 & 0.04616 & 0.05787 & 0.07455 & 0.09862 & 0.1336 & 0.1847 & 0.2590 & 0.3667 & 0.5215 \\
0.04 & 0.04457 & 0.05628 & 0.07295 & 0.09703 & 0.1320 & 0.1831 & 0.2574 & 0.3651 & 0.5199 \\
0.05 & 0.04298 & 0.05468 & 0.07136 & 0.09544 & 0.1305 & 0.1815 & 0.2558 & 0.3635 & 0.5183 \\
\hline

\multicolumn{10}{c}{$P_c$} \\
\hline
0.01 & 0.005380 & 0.007880 & 0.01186 & 0.01820 & 0.02833 & 0.04440 & 0.06969 & 0.1091 & 0.1697 \\
\hline

\multicolumn{10}{c}{$P_c v_c / T_c$} \\
\hline
0.01 & 0.4307 & 0.4774 & 0.5285 & 0.5810 & 0.6316 & 0.6776 & 0.7173 & 0.7504 & 0.7770 \\
0.02 & 0.4451 & 0.4901 & 0.5395 & 0.5902 & 0.6390 & 0.6834 & 0.7217 & 0.7536 & 0.7794 \\
0.03 & 0.4604 & 0.5036 & 0.5511 & 0.5998 & 0.6467 & 0.6893 & 0.7261 & 0.7569 & 0.7818 \\
0.04 & 0.4769 & 0.5179 & 0.5631 & 0.6096 & 0.6545 & 0.6953 & 0.7306 & 0.7602 & 0.7842 \\
0.05 & 0.4945 & 0.5329 & 0.5756 & 0.6198 & 0.6624 & 0.7013 & 0.7352 & 0.7635 & 0.7866 \\
\hline
\end{tabular}
\caption{Critical quantities for varying $\lambda$ and $N$. Here $a=0.01 \times 10^3,\,Q=1,\,\alpha=0.10$.}
\label{tab:1}
\end{table*}

\begin{table*}[tbhp]
\centering
\small
\begin{tabular}{|c|ccccccccc|}
\hline
$N \backslash \lambda$
& 0.1 & 0.2 & 0.3 & 0.4 & 0.5 & 0.6 & 0.7 & 0.8 & 0.9 \\
\hline

\multicolumn{10}{c}{$v_c$} \\
\hline
0.01 & 4.054 & 3.790 & 3.545 & 3.319 & 3.110 & 2.919 & 2.743 & 2.583 & 2.435 \\
\hline

\multicolumn{10}{c}{$T_c$} \\
\hline
0.01 & 0.0441 & 0.0545 & 0.0693 & 0.0909 & 0.1224 & 0.1688 & 0.2367 & 0.3357 & 0.4791 \\
0.02 & 0.0425 & 0.0529 & 0.0677 & 0.0893 & 0.1209 & 0.1672 & 0.2351 & 0.3341 & 0.4775 \\
0.03 & 0.0409 & 0.0513 & 0.0661 & 0.0877 & 0.1193 & 0.1656 & 0.2335 & 0.3325 & 0.4760 \\
0.04 & 0.0393 & 0.0497 & 0.0645 & 0.0861 & 0.1177 & 0.1640 & 0.2319 & 0.3309 & 0.4744 \\
0.05 & 0.0377 & 0.0481 & 0.0630 & 0.0845 & 0.1161 & 0.1624 & 0.2303 & 0.3293 & 0.4728 \\
\hline

\multicolumn{10}{c}{$P_c$} \\
\hline
0.01 & 0.004558 & 0.006687 & 0.01010 & 0.01562 & 0.02450 & 0.03874 & 0.06138 & 0.09698 & 0.1523 \\
\hline

\multicolumn{10}{c}{$P_c v_c / T_c$} \\
\hline
0.01 & 0.4195 & 0.4655 & 0.5168 & 0.5703 & 0.6224 & 0.6701 & 0.7116 & 0.7461 & 0.7739 \\
0.02 & 0.4352 & 0.4795 & 0.5289 & 0.5804 & 0.6306 & 0.6765 & 0.7164 & 0.7496 & 0.7765 \\
0.03 & 0.4522 & 0.4943 & 0.5416 & 0.5910 & 0.6390 & 0.6830 & 0.7213 & 0.7532 & 0.7791 \\
0.04 & 0.4705 & 0.5102 & 0.5550 & 0.6019 & 0.6476 & 0.6896 & 0.7262 & 0.7568 & 0.7817 \\
0.05 & 0.4904 & 0.5271 & 0.5690 & 0.6132 & 0.6565 & 0.6964 & 0.7312 & 0.7605 & 0.7844 \\
\hline
\end{tabular}
\caption{Critical quantities for varying $\lambda$ and $N$. Here $a=0.01 \times 10^3,\,Q=1,\,\alpha=0.15$.}
\label{tab:2}
\end{table*}

A particularly instructive limit is obtained by setting $c_5=0$, which implies $a=0$. This corresponds to the absence of the Euler--Heisenberg nonlinear parameter, and the spacetime then reduces to the charged Letelier-AdS black hole surrounded by quintessence and PFDM. In that case, the critical volume can be written explicitly as
\begin{equation}
    v_c=\frac{-3c_3+\sqrt{9c_3^2+24c_2c_4}}{2c_2}.
    \label{cc7}
\end{equation}
Substituting $v_c$ into Eqs.~(\ref{cc5}) and (\ref{cc6}), one can obtain the corresponding critical temperature and pressure, respectively.

From a physical point of view, the main outcome of the analysis is that each external sector deforms the thermodynamic phase structure in a distinguishable manner. The string cloud parameter modifies the effective geometric background; the PFDM sector introduces a logarithmic-type long-range correction; quintessence changes the large-distance behavior through its equation of state; and the Euler--Heisenberg term primarily affects the strong-field region. As a result, the location of the critical point and the thermal stability of the black hole are controlled by the combined action of short-range nonlinear electromagnetic effects and long-range matter distributions.

\section{Sparsity of Hawking Radiation}\label{sec:5}

In addition to equilibrium thermodynamic quantities, it is also instructive to examine the dynamical character of the Hawking emission process \cite{Hawking1975}. A useful diagnostic in this context is the sparsity parameter, which quantifies the temporal separation between successive Hawking quanta. A large value of the sparsity parameter indicates an intermittent Hawking flux, whereas smaller values correspond to a more continuous emission spectrum \cite{Visser2017Sparsity}. The sparsity of Hawking radiation have recently been studied in a variety of black hole space-times, including those arising in nonlinear electrodynamics and AdS geometries \cite{AhmedPDU2025,AhmedBTZ2025,AbbasAli2023,Ahmed2026ModMaxAdS,Ahmed2026SkyrmionBH,Ahmed2026KR_PFDM,Ahmed2026BumblebeeMonopole,Ahmed2026BumblebeeTsallis}.

In the present context, the sparsity parameter is indirectly influenced by the Hawking temperature and the effective emitting area. Therefore, any modification in the horizon geometry or in the temperature profile caused by nonlinear electrodynamics, quintessence, PFDM, or the cloud of strings will also affect the radiation pattern. This makes sparsity an interesting complementary observable that connects the thermodynamic sector to the phenomenology of black hole evaporation.

The sparsity of radiation is characterized by the dimensionless parameter \cite{Gray2016,Page1976}
\begin{equation}
    \eta=\frac{\mathcal{C}}{\tilde{g}}\,\frac{\lambda_t^2}{\mathcal{A}_{\rm eff}},
    \label{dd1}
\end{equation}
where $\mathcal{C}$ is a constant, $\tilde{g}$ is the number of photon degrees of freedom, $\mathcal{A}_{\rm eff}=\frac{27}{4}A_{\rm BH}$ is the effective horizon area, $A_{\rm BH}$ is the horizon area, and $\lambda_t=\frac{2\pi}{T_H}$ is the thermal wavelength.

Using the above ingredients, the sparsity parameter for $\omega_q=-2/3$ becomes
\begin{align}
\eta&=\frac{64\pi^3}{27[\Delta(r_h)]^{2}},\nonumber\\
\Delta(r_h)&=1-\alpha-\frac{Q^2}{r_h^2}
+\frac{aQ^4}{4r_h^6}-2Nr_h+\frac{\lambda}{r_h}+\frac{3r_h^2}{\ell^2}.\label{dd4}
\end{align}
\begin{figure}[tbhp]
\centering
\includegraphics[width=\linewidth]{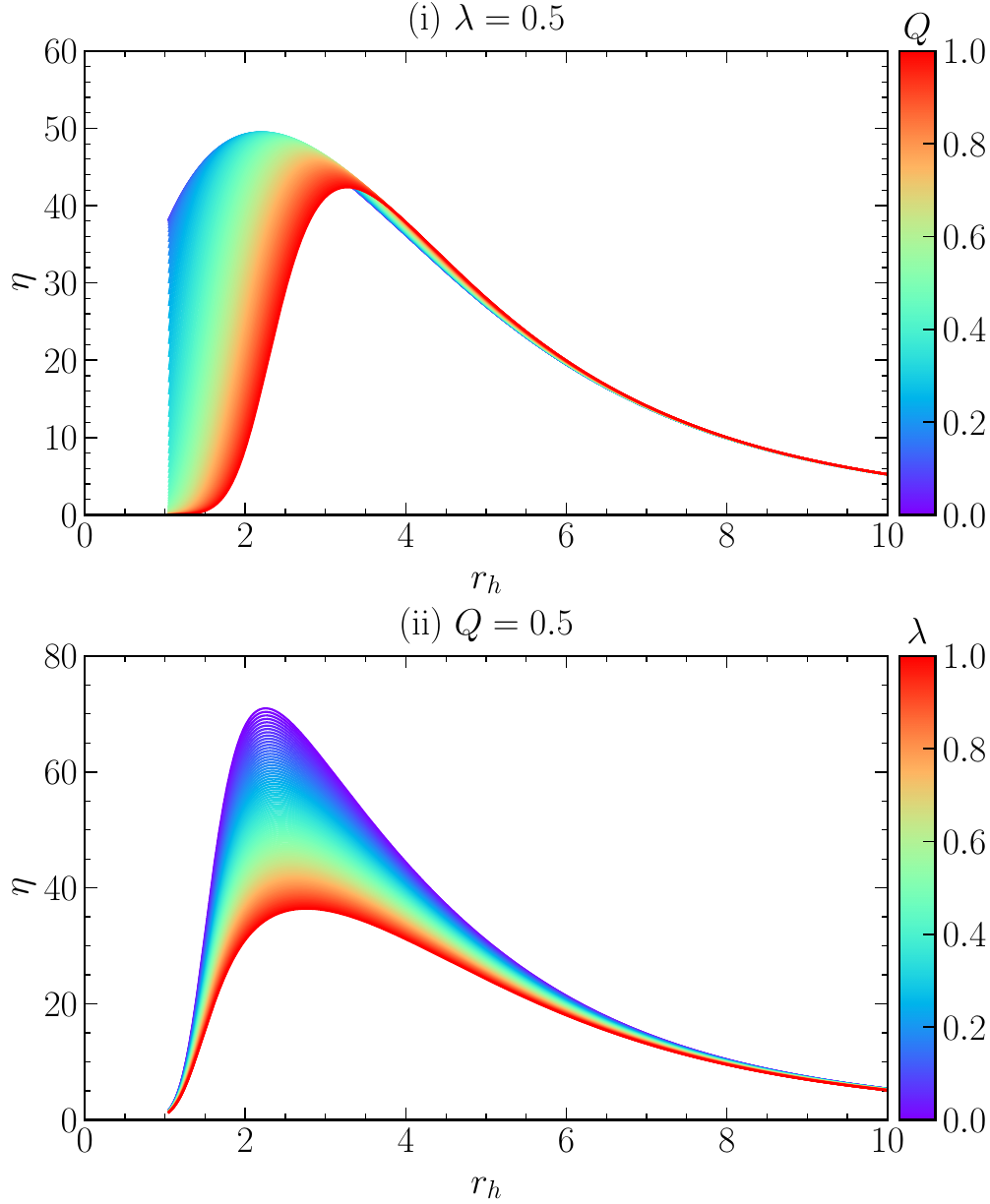}
\caption{Two-panel behavior of the sparsity parameter $\eta$ as a function of the horizon radius $r_h$. In the upper panel, $\lambda=0.5$ is fixed, and the charge $Q$ is varied; in the lower panel, $Q=0.5$ is fixed, and $\lambda$ is varied. The remaining parameters are $\alpha=0.1$, $P=\tfrac{0.03}{8\pi}$, $N=0.01$, $\omega_q=-2/3$, and $a=0.50\times 10^3$.}
\label{fig:eta}
\end{figure}
\begin{figure}[tbhp]
\centering
\includegraphics[width=0.82\linewidth]{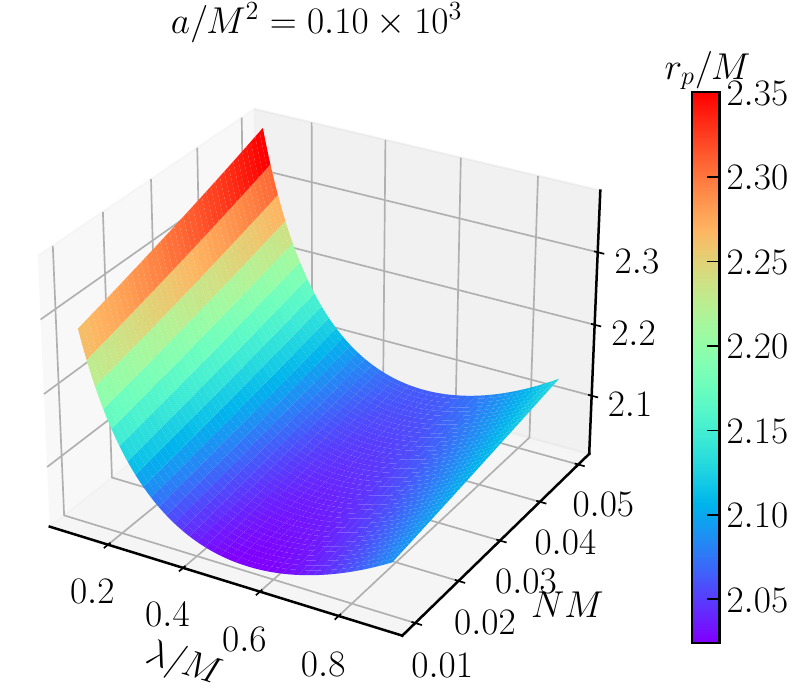}\\[0.8em]
\includegraphics[width=0.82\linewidth]{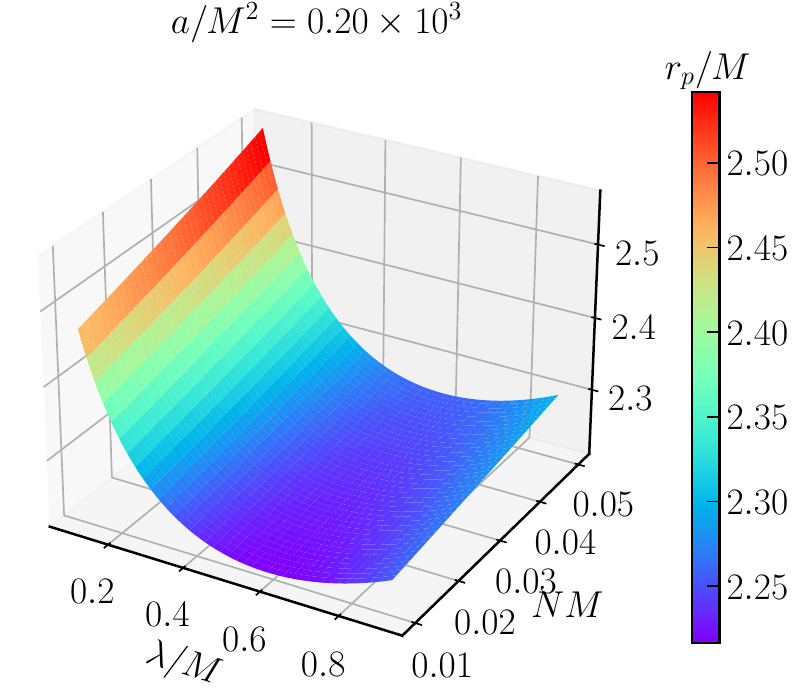}
\caption{Photon-sphere radius $r_p/M$ as a function of the quintessence parameter $N$ and the PFDM parameter $\lambda$ for $a/M^2=0.10\times 10^3$ (upper panel) and $a/M^2=0.20\times 10^3$ (lower panel). Here $Q/M=1$ and $\alpha=0.1$.}
\label{fig:photon}
\end{figure}
\begin{figure}[tbhp]
\centering
\includegraphics[width=0.82\linewidth]{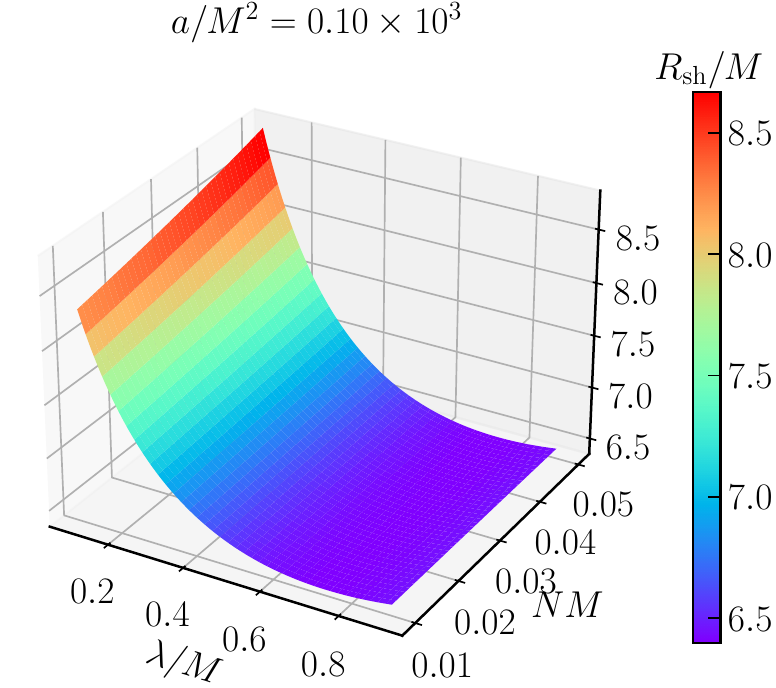}\\[0.8em]
\includegraphics[width=0.82\linewidth]{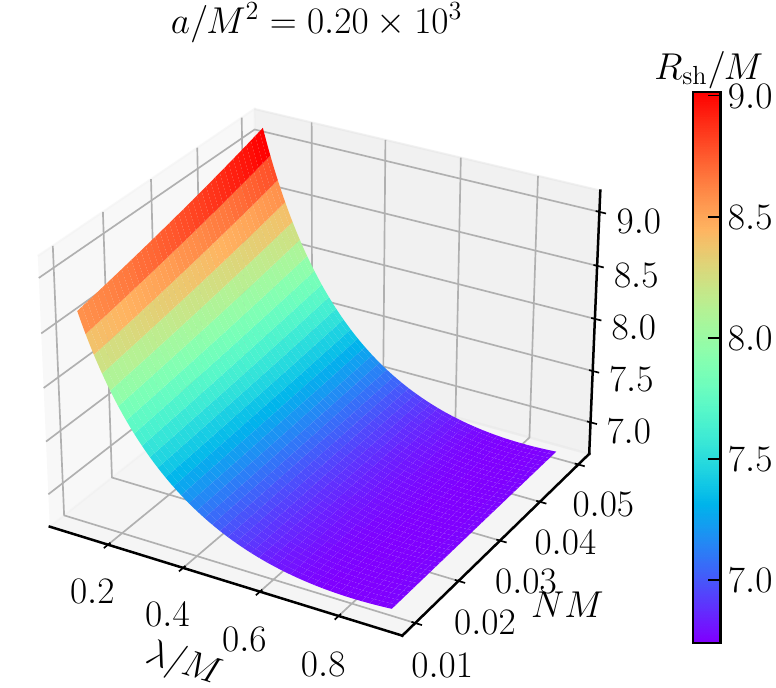}
\caption{Shadow radius $R_{\rm sh}/M$ as a function of the quintessence parameter $N$ and the PFDM parameter $\lambda$ for $a/M^2=0.10\times 10^3$ (upper panel) and $a/M^2=0.20\times 10^3$ (lower panel). Here $Q/M=1$, $\alpha=0.1$, $r_O/M=50$, and $\ell/M=10$.}
\label{fig:shadow}
\end{figure}

The resulting expression shows that the sparsity is controlled by the same combination of parameters that governs the Hawking temperature. Consequently, whenever the temperature is suppressed, the radiation becomes sparser. From this point of view, the charge and the string-cloud contribution tend to enhance the intermittency of the Hawking flux, while the pressure term generally works in the opposite direction by increasing the temperature. The Euler--Heisenberg correction is especially relevant in the small-horizon regime, where it can noticeably change the emission profile. The quintessence and PFDM sectors also contribute nontrivially, showing that environmental matter not only alters the equilibrium thermodynamics but also leaves an imprint on the dynamical character of Hawking radiation. Figure~\ref{fig:eta} shows that the sparsity develops a pronounced peak in the small-$r_h$ regime and then decays for larger radius, while both $Q$ and $\lambda$ substantially modify the peak height and the relaxation profile.

\section{Photon Sphere and Black Hole Shadow}\label{sec:6}

In this section, we study the photon sphere and the black hole shadow and analyze how external matter fields and the nonlinear electrodynamics sector modify these optical observables. This problem is especially timely in view of the Event Horizon Telescope observations and the large recent literature on optical signatures of matter-surrounded or dark-matter-dressed black holes \cite{Akiyama2019L1,Akiyama2019L4,Akiyama2022L12,PerlickTsupko2022,HouXuWang2018,HaroonEtAl2019,HeydariFardEtAl2023,AhmedHernquist2025}. The equation of motion for photon particles is obtained as
\begin{equation}
    \dot r^2+V_{\rm eff}=\mathrm{E}^2,\label{tt1}
\end{equation}
where $V_{\rm eff}$ is the effective potential governing the photon dynamics. It is given by
\begin{equation}
    V_{\rm eff}=\frac{\mathrm{L}^2}{r^2}\,f(r).\label{tt2}
\end{equation}
Circular null orbits satisfy the condition $\dot r=0$ together with
\begin{equation}
    \frac{\partial }{\partial r}\left(\frac{f(r)}{r^2}\right)=0.\label{tt3}
\end{equation}
This relation implies 
\begin{equation}
    f(r_p)-r_p\,f'(r_p)/2=0.\label{photon}
\end{equation}
Substituting the metric function $f(r)$ into the above relation yields a nonlinear equation for the photon-sphere radius. Because of the logarithmic PFDM term, a closed analytical solution is not available in general, so $r_p$ must be determined numerically for representative parameter choices.

Moreover, the spacetime is not asymptotically flat, since at large radius the lapse function behaves as
\begin{equation}
    \lim_{ r \to \infty} f(r)=1-\alpha+\frac{r^2}{\ell^2}.
\end{equation}
Following the approach in \cite{PerlickTsupko2022}, we define
\begin{align}
\mathcal{F}(r)\equiv -\frac{2M}{r}+\mathcal{H}(r),
\label{aa0e}
\end{align}
so that, for the state parameter $\omega_q=-2/3$, the shadow radius can be written as \cite{PerlickTsupko2022}
\begin{align}
R_{\rm sh}
=r_p\,\sqrt{\frac{\mathcal{F}(r_O)}{\mathcal{F}(r_p)}}.
    \label{shadow}
\end{align}
Here, $\mathcal{F}(r_O)$ and $\mathcal{F}(r_p)$ denote the metric function
evaluated at the observer position $r=r_O$ and at the photon-sphere radius
$r=r_p$, respectively.

\begin{table*}[tbhp]
\centering
\begin{tabular}{|c|ccccccccc|}
\hline
$N\,M$ $\backslash$ $\lambda/M$ 
& 0.10 & 0.20 & 0.30 & 0.40 & 0.50 & 0.60 & 0.70 & 0.80 & 0.90 \\
\hline
0.01 & 2.28188 & 2.13807 & 2.06963 & 2.03662 & 2.02440 & 2.02629 & 2.03887 & 2.06020 & 2.08913 \\
0.02 & 2.30308 & 2.15246 & 2.08113 & 2.04672 & 2.03382 & 2.03545 & 2.04803 & 2.06958 & 2.09890 \\
0.03 & 2.32570 & 2.16760 & 2.09313 & 2.05722 & 2.04358 & 2.04492 & 2.05752 & 2.07929 & 2.10901 \\
0.04 & 2.34990 & 2.18357 & 2.10569 & 2.06814 & 2.05371 & 2.05474 & 2.06733 & 2.08934 & 2.11947 \\
0.05 & 2.37591 & 2.20044 & 2.11884 & 2.07953 & 2.06425 & 2.06493 & 2.07751 & 2.09975 & 2.13031 \\
\hline
\end{tabular}
\caption{Numerical values of the photon-sphere radius $r_p/M$ for varying $\lambda$ and $N$. Here $a/M^2=0.10 \times 10^3$, $Q/M=1$, and $\alpha=0.1$.}
\label{tab:3}
\end{table*}

\begin{table*}[tbhp]
\centering
\begin{tabular}{|c|ccccccccc|}
\hline
$N\,M$ $\backslash$ $\lambda/M$ 
& 0.10 & 0.20 & 0.30 & 0.40 & 0.50 & 0.60 & 0.70 & 0.80 & 0.90 \\
\hline
0.00 & 2.45130 & 2.32506 & 2.26036 & 2.22604 & 2.21007 & 2.20670 & 2.21285 & 2.22674 & 2.24728 \\
0.01 & 2.47125 & 2.33943 & 2.27217 & 2.23651 & 2.21982 & 2.21611 & 2.22217 & 2.23615 & 2.25693 \\
0.02 & 2.49242 & 2.35451 & 2.28448 & 2.24739 & 2.22993 & 2.22585 & 2.23180 & 2.24587 & 2.26690 \\
0.03 & 2.51496 & 2.37036 & 2.29733 & 2.25869 & 2.24041 & 2.23593 & 2.24177 & 2.25593 & 2.27721 \\
0.04 & 2.53905 & 2.38707 & 2.31078 & 2.27046 & 2.25130 & 2.24638 & 2.25208 & 2.26633 & 2.28788 \\
0.05 & 2.56487 & 2.40472 & 2.32486 & 2.28274 & 2.26261 & 2.25722 & 2.26278 & 2.27711 & 2.29893 \\
\hline
\end{tabular}
\caption{Numerical values of the photon-sphere radius $r_p/M$ for varying $\lambda$ and $N$. Here $a/M^2=0.20 \times 10^3$, $Q/M=1$, and $\alpha=0.1$.}
\label{tab:4}
\end{table*}

\begin{table*}[tbhp]
\centering
\begin{tabular}{|c|ccccccccc|}
\hline
$N\,M$ $\backslash$ $\lambda/M$ 
& 0.10 & 0.20 & 0.30 & 0.40 & 0.50 & 0.60 & 0.70 & 0.80 & 0.90 \\
\hline
0.01 & 8.34065 & 7.50818 & 7.01951 & 6.72116 & 6.54203 & 6.44413 & 6.40515 & 6.41094 & 6.45196 \\
0.02 & 8.43982 & 7.56174 & 7.04911 & 6.73716 & 6.54999 & 6.44734 & 6.40569 & 6.41027 & 6.45111 \\
0.03 & 8.54948 & 7.62001 & 7.08079 & 6.75391 & 6.55797 & 6.45015 & 6.40561 & 6.40885 & 6.44948 \\
0.04 & 8.67160 & 7.68376 & 7.11484 & 6.77148 & 6.56594 & 6.45248 & 6.40480 & 6.40657 & 6.44695 \\
0.05 & 8.80870 & 7.75394 & 7.15162 & 6.78997 & 6.57387 & 6.45424 & 6.40312 & 6.40329 & 6.44335 \\
\hline
\end{tabular}
\caption{Numerical values of shadow radius $R_{sh}/M$ for varying $\lambda$ and $N$. Here $a/M^2=0.10 \times 10^3,\,Q/M=1,\,\alpha=0.1,\,r_0=50/M,\,\ell/M=10$.}
\label{tab:5}
\end{table*}

\begin{table*}[tbhp]
\centering
\begin{tabular}{|c|ccccccccc|}
\hline
$N\,M$ $\backslash$ $\lambda/M$ 
& 0.10 & 0.20 & 0.30 & 0.40 & 0.50 & 0.60 & 0.70 & 0.80 & 0.90 \\
\hline
0.01 & 8.66314 & 7.89277 & 7.42383 & 7.12453 & 6.93357 & 6.81767 & 6.75695 & 6.73868 & 6.75423 \\
0.02 & 8.76507 & 7.95147 & 7.45891 & 7.14562 & 6.94601 & 6.82471 & 6.76070 & 6.74063 & 6.75549 \\
0.03 & 8.87796 & 8.01553 & 7.49667 & 7.16795 & 6.95887 & 6.83167 & 6.76411 & 6.74208 & 6.75616 \\
0.04 & 9.00391 & 8.08585 & 7.53750 & 7.19168 & 6.97219 & 6.83855 & 6.76709 & 6.74293 & 6.75614 \\
0.05 & 9.14557 & 8.16353 & 7.58188 & 7.21700 & 6.98600 & 6.84528 & 6.76956 & 6.74305 & 6.75531 \\
\hline
\end{tabular}
\caption{Numerical values of shadow radius $R_{sh}/M$ for varying $\lambda$ and $N$. Here $a/M^2=0.20 \times 10^3,\,Q/M=1,\,\alpha=0.1,\,r_0=50/M,\,\ell/M=10$.}
\label{tab:6}
\end{table*}

In Tables \ref{tab:3}--\ref{tab:4}, we present the numerical values of the photon sphere radius by varying the quintessence parameter $N$ and the PFDM parameter $\lambda$, for the state parameter $w=-2/3$, considering two different values of the EH parameter $a$.

Similarly, in Tables \ref{tab:5}--\ref{tab:6}, we report the corresponding numerical values of the shadow radius under the same variations of $N$ and $\lambda$, for $w=-2/3$ and for two distinct values of the EH parameter $a$.

Furthermore, Figures \ref{fig:photon} and \ref{fig:shadow} display the graphical behavior of the photon-sphere and shadow radii, respectively, for $w=-2/3$ and for the two chosen values of the EH parameter $a$. The surfaces show that, for fixed $a$, increasing the quintessence intensity $N$ generally enlarges both $r_p$ and $R_{\rm sh}$, whereas the dependence on the PFDM parameter $\lambda$ is mildly non-monotonic: both observables decrease from small $\lambda$, reach a shallow minimum, and then increase slightly again at larger $\lambda$. The larger Euler--Heisenberg coupling shifts both radii upward overall, confirming that nonlinear electrodynamics leaves a measurable imprint on the optical sector.

\section{Energy Emission Rate}\label{sec:7}

The production of particle pairs near the event horizon gives rise to Hawking radiation, originating from quantum fluctuations in the black hole's background. Black hole evaporation proceeds as positive-energy particles escape from the near-horizon region via a quantum tunneling process, while their negative-energy counterparts are absorbed by the black hole.

In this work, we investigate the energy emission rate associated with the considered black hole geometry within the framework of EH theory surrounded by PFDM and quintessence in the presence of a cloud of strings. In the high-energy regime, the absorption cross-section typically exhibits oscillatory behavior around a limiting constant value $\sigma_{\rm lim}$.  

In the geometric-optics regime, this absorption cross-section approaches the limiting value \cite{WeiLiu2013}
\begin{equation}
\sigma_{\rm lim}\approx \pi R_{\rm sh}^2,\label{ee1-new}
\end{equation}
where $R_{\rm sh}$ is given in (\ref{shadow}). 

Within this approximation, the spectral energy emission rate is given by \cite{WeiLiu2013,DecaniniEtAl2011,Mashhoon1973,AhmedPDU2025,Sanchez1978,ShahzadEtAl2025}
\begin{equation}
\frac{d^2\mathbb{E}}{d\omega\,dt}=\frac{2\pi^2\sigma_{\rm lim}}{e^{\omega/T_H}-1}\,\omega^3,\label{ee2-new}
\end{equation}
where $\omega$ denotes the emitted frequency and $T_H$ is the Hawking temperature. Using Eqs.~(\ref{ee1-new}) and (\ref{ee2-new}), the emission rate can be written in the compact form
\begin{equation}
\frac{d^2\mathbb{E}}{d\omega\,dt}=\frac{2\pi^3 R^2_{\rm sh}\,\omega^3}{e^{\omega/T_H}-1}
=\frac{2\pi^3 R^2_{\rm sh}\,\omega^3}{\exp\!\left(\frac{4\pi r_h\,\omega}{\Delta(r_h)}\right)-1}.\label{ee3-new}
\end{equation}

This expression shows that the emission spectrum is controlled by two competing effects: the shadow radius sets the overall effective capture area, whereas the Hawking temperature governs the exponential suppression at high frequency. Therefore, any parameter that lowers $T_H$ suppresses the high-frequency tail and shifts the dominant emission toward smaller $\omega$, while an increase in $R_{\rm sh}$ enhances the overall amplitude. In the present geometry, the Euler--Heisenberg term and the surrounding matter fields influence both quantities simultaneously, so the radiative signal carries information about both the near-horizon electromagnetic sector and the external environment.

\section{Conclusion}\label{sec:8}

In this paper, we have investigated the thermodynamic, radiative, and optical properties of an Euler--Heisenberg AdS black hole surrounded by quintessence, perfect fluid dark matter, and a cloud of strings. Starting from the corresponding lapse function, we derived the relevant thermodynamic quantities in the extended phase-space formalism and verified that the generalized first law and Smarr relation remain consistent when the PFDM and quintessence sectors are treated as genuine thermodynamic variables.

We have shown that the metric function, horizon structure, and thermodynamic response of the black hole are all significantly affected by the cloud-of-strings parameter, the quintessence intensity, the PFDM parameter, the electric charge, and the Euler--Heisenberg coupling. In particular, the Hawking temperature develops a nontrivial profile with a local maximum, while the heat capacity exhibits divergences that separate thermodynamically stable and unstable branches. These results indicate that the surrounding matter fields do not merely deform the background geometry but actively reorganize the system's thermal stability.

We also analyzed the equation of state and the associated criticality conditions. The resulting $P$--$v$ isotherms confirm a Van der Waals--type phase structure, with a subcritical oscillatory branch, a critical inflection point, and monotonic supercritical behavior. Since the Euler--Heisenberg term enters through a higher-order inverse-power correction, its effect is especially pronounced in the small-volume regime, where it can substantially shift the critical point.

The sparsity analysis further shows that the intermittency of Hawking radiation is governed by the same combination of parameters that controls the temperature. The sparsity parameter develops a pronounced peak at small horizon radius and then decreases as the horizon grows, demonstrating that nonlinear electrodynamics and environmental matter affect not only equilibrium thermodynamics but also the temporal structure of the Hawking flux.

Finally, the photon-sphere and shadow analyses reveal a complementary optical signature of the model. The numerical tables and 3D surfaces show that increasing the quintessence intensity generally increases both the photon-sphere radius and the shadow radius, whereas the PFDM parameter exhibits a shallow, non-monotonic behavior. The Euler--Heisenberg coupling shifts these optical observables upward overall, and the same ingredients enter the geometric-optics emission rate through $R_{\rm sh}$ and $T_H$.

Overall, our results show that nonlinear electrodynamics, together with surrounding dark-sector matter and a cloud of strings, leaves correlated geometric, thermodynamic, radiative, and optical imprints on AdS black holes. Possible extensions of this work include a full numerical study of the spectral energy-emission curves, greybody factors beyond the geometric-optics approximation, quasinormal modes, and Joule--Thomson expansion in the same background, as well as a broader comparison with other dark-matter profiles and nonlinear-electrodynamics models.

\scriptsize

\section*{Acknowledgments}

F.A. acknowledges the Inter University Centre for Astronomy and Astrophysics (IUCAA), Pune, India for granting visiting associateship.  E. O. Silva acknowledges the support from Conselho Nacional de Desenvolvimento Cient\'{i}fico e Tecnol\'{o}gico (CNPq) (grants 306308/2022-3), Funda\c c\~ao de Amparo \`{a} Pesquisa e ao Desenvolvimento Cient\'{i}fico e Tecnol\'{o}gico do Maranh\~ao (FAPEMA) (grants UNIVERSAL-06395/22), and Coordena\c c\~ao de Aperfei\c coamento de Pessoal de N\'{i}vel Superior (CAPES) - Brazil (Code 001).

%


\begin{thebibliography}{55}%
	\makeatletter
	\providecommand \@ifxundefined [1]{%
		\@ifx{#1\undefined}
	}%
	\providecommand \@ifnum [1]{%
		\ifnum #1\expandafter \@firstoftwo
		\else \expandafter \@secondoftwo
		\fi
	}%
	\providecommand \@ifx [1]{%
		\ifx #1\expandafter \@firstoftwo
		\else \expandafter \@secondoftwo
		\fi
	}%
	\providecommand \natexlab [1]{#1}%
	\providecommand \enquote  [1]{``#1''}%
	\providecommand \bibnamefont  [1]{#1}%
	\providecommand \bibfnamefont [1]{#1}%
	\providecommand \citenamefont [1]{#1}%
	\providecommand \href@noop [0]{\@secondoftwo}%
	\providecommand \href [0]{\begingroup \@sanitize@url \@href}%
	\providecommand \@href[1]{\@@startlink{#1}\@@href}%
	\providecommand \@@href[1]{\endgroup#1\@@endlink}%
	\providecommand \@sanitize@url [0]{\catcode `\\12\catcode `\$12\catcode
		`\&12\catcode `\#12\catcode `\^12\catcode `\_12\catcode `\%12\relax}%
	\providecommand \@@startlink[1]{}%
	\providecommand \@@endlink[0]{}%
	\providecommand \url  [0]{\begingroup\@sanitize@url \@url }%
	\providecommand \@url [1]{\endgroup\@href {#1}{\urlprefix }}%
	\providecommand \urlprefix  [0]{URL }%
	\providecommand \Eprint [0]{\href }%
	\providecommand \doibase [0]{https://doi.org/}%
	\providecommand \selectlanguage [0]{\@gobble}%
	\providecommand \bibinfo  [0]{\@secondoftwo}%
	\providecommand \bibfield  [0]{\@secondoftwo}%
	\providecommand \translation [1]{[#1]}%
	\providecommand \BibitemOpen [0]{}%
	\providecommand \bibitemStop [0]{}%
	\providecommand \bibitemNoStop [0]{.\EOS\space}%
	\providecommand \EOS [0]{\spacefactor3000\relax}%
	\providecommand \BibitemShut  [1]{\csname bibitem#1\endcsname}%
	\let\auto@bib@innerbib\@empty
	\bibitem [{\citenamefont {Bekenstein}(1973)}]{Bekenstein1973}%
	\BibitemOpen
	\bibfield  {author} {\bibinfo {author} {\bibfnamefont {J.~D.}\ \bibnamefont
			{Bekenstein}},\ }\href {https://doi.org/10.1103/PhysRevD.7.2333} {\bibfield
		{journal} {\bibinfo  {journal} {Phys. Rev. D}\ }\textbf {\bibinfo {volume}
			{7}},\ \bibinfo {pages} {2333} (\bibinfo {year} {1973})}\BibitemShut
	{NoStop}%
	\bibitem [{\citenamefont {Bardeen}\ \emph {et~al.}(1973)\citenamefont
		{Bardeen}, \citenamefont {Carter},\ and\ \citenamefont
		{Hawking}}]{BardeenCarterHawking1973}%
	\BibitemOpen
	\bibfield  {author} {\bibinfo {author} {\bibfnamefont {J.~M.}\ \bibnamefont
			{Bardeen}}, \bibinfo {author} {\bibfnamefont {B.}~\bibnamefont {Carter}},\
		and\ \bibinfo {author} {\bibfnamefont {S.~W.}\ \bibnamefont {Hawking}},\
	}\href {https://doi.org/10.1007/BF01645742} {\bibfield  {journal} {\bibinfo
			{journal} {Commun. Math. Phys.}\ }\textbf {\bibinfo {volume} {31}},\ \bibinfo
		{pages} {161} (\bibinfo {year} {1973})}\BibitemShut {NoStop}%
	\bibitem [{\citenamefont {Hawking}(1975)}]{Hawking1975}%
	\BibitemOpen
	\bibfield  {author} {\bibinfo {author} {\bibfnamefont {S.~W.}\ \bibnamefont
			{Hawking}},\ }\href {https://doi.org/10.1007/BF02345020} {\bibfield
		{journal} {\bibinfo  {journal} {Commun. Math. Phys.}\ }\textbf {\bibinfo
			{volume} {43}},\ \bibinfo {pages} {199} (\bibinfo {year} {1975})}\BibitemShut
	{NoStop}%
	\bibitem [{\citenamefont {Hawking}\ and\ \citenamefont
		{Page}(1983)}]{HawkingPage1983}%
	\BibitemOpen
	\bibfield  {author} {\bibinfo {author} {\bibfnamefont {S.~W.}\ \bibnamefont
			{Hawking}}\ and\ \bibinfo {author} {\bibfnamefont {D.~N.}\ \bibnamefont
			{Page}},\ }\href {https://doi.org/10.1007/BF01208266} {\bibfield  {journal}
		{\bibinfo  {journal} {Commun. Math. Phys.}\ }\textbf {\bibinfo {volume}
			{87}},\ \bibinfo {pages} {577} (\bibinfo {year} {1983})}\BibitemShut
	{NoStop}%
	\bibitem [{\citenamefont {Kastor}\ \emph {et~al.}(2009)\citenamefont {Kastor},
		\citenamefont {Ray},\ and\ \citenamefont {Traschen}}]{KastorEtAl2009}%
	\BibitemOpen
	\bibfield  {author} {\bibinfo {author} {\bibfnamefont {D.}~\bibnamefont
			{Kastor}}, \bibinfo {author} {\bibfnamefont {S.}~\bibnamefont {Ray}},\ and\
		\bibinfo {author} {\bibfnamefont {J.}~\bibnamefont {Traschen}},\ }\href
	{https://doi.org/10.1088/0264-9381/26/19/195011} {\bibfield  {journal}
		{\bibinfo  {journal} {Class. Quantum Grav.}\ }\textbf {\bibinfo {volume}
			{26}},\ \bibinfo {pages} {195011} (\bibinfo {year} {2009})}\BibitemShut
	{NoStop}%
	\bibitem [{\citenamefont {Kubiz\v{n}\'ak}\ and\ \citenamefont
		{Mann}(2012)}]{KubiznakMann2012}%
	\BibitemOpen
	\bibfield  {author} {\bibinfo {author} {\bibfnamefont {D.}~\bibnamefont
			{Kubiz\v{n}\'ak}}\ and\ \bibinfo {author} {\bibfnamefont {R.~B.}\
			\bibnamefont {Mann}},\ }\href {https://doi.org/10.1007/JHEP07(2012)033}
	{\bibfield  {journal} {\bibinfo  {journal} {JHEP}\ }\textbf {\bibinfo
			{volume} {2012}}\bibinfo  {number} { (7)},\ \bibinfo {pages}
		{033}}\BibitemShut {NoStop}%
	\bibitem [{\citenamefont {Gunasekaran}\ \emph {et~al.}(2012)\citenamefont
		{Gunasekaran}, \citenamefont {Kubiz\v{n}\'ak},\ and\ \citenamefont
		{Mann}}]{GunasekaranEtAl2012}%
	\BibitemOpen
	\bibfield  {number} {  }\bibfield  {author} {\bibinfo {author} {\bibfnamefont
			{S.}~\bibnamefont {Gunasekaran}}, \bibinfo {author} {\bibfnamefont
			{D.}~\bibnamefont {Kubiz\v{n}\'ak}},\ and\ \bibinfo {author} {\bibfnamefont
			{R.~B.}\ \bibnamefont {Mann}},\ }\href
	{https://doi.org/10.1007/JHEP11(2012)110} {\bibfield  {journal} {\bibinfo
			{journal} {JHEP}\ }\textbf {\bibinfo {volume} {2012}}\bibinfo  {number} {
			(11)},\ \bibinfo {pages} {110}}\BibitemShut {NoStop}%
	\bibitem [{\citenamefont {Kubiz\v{n}\'ak}\ and\ \citenamefont
		{Mann}(2015)}]{KubiznakMann2015}%
	\BibitemOpen
	\bibfield  {number} {  }\bibfield  {author} {\bibinfo {author} {\bibfnamefont
			{D.}~\bibnamefont {Kubiz\v{n}\'ak}}\ and\ \bibinfo {author} {\bibfnamefont
			{R.~B.}\ \bibnamefont {Mann}},\ }\href
	{https://doi.org/10.1139/CJP-2014-0465} {\bibfield  {journal} {\bibinfo
			{journal} {Can. J. Phys.}\ }\textbf {\bibinfo {volume} {93}},\ \bibinfo
		{pages} {999} (\bibinfo {year} {2015})}\BibitemShut {NoStop}%
	\bibitem [{\citenamefont {Kubiz\v{n}\'ak}\ \emph {et~al.}(2017)\citenamefont
		{Kubiz\v{n}\'ak}, \citenamefont {Mann},\ and\ \citenamefont
		{Teo}}]{KubiznakMannTeo2017}%
	\BibitemOpen
	\bibfield  {author} {\bibinfo {author} {\bibfnamefont {D.}~\bibnamefont
			{Kubiz\v{n}\'ak}}, \bibinfo {author} {\bibfnamefont {R.~B.}\ \bibnamefont
			{Mann}},\ and\ \bibinfo {author} {\bibfnamefont {M.}~\bibnamefont {Teo}},\
	}\href {https://doi.org/10.1088/1361-6382/aa5c69} {\bibfield  {journal}
		{\bibinfo  {journal} {Class. Quantum Grav.}\ }\textbf {\bibinfo {volume}
			{34}},\ \bibinfo {pages} {063001} (\bibinfo {year} {2017})}\BibitemShut
	{NoStop}%
	\bibitem [{\citenamefont {Heisenberg}\ and\ \citenamefont
		{Euler}(1936)}]{HeisenbergEuler1936}%
	\BibitemOpen
	\bibfield  {author} {\bibinfo {author} {\bibfnamefont {W.}~\bibnamefont
			{Heisenberg}}\ and\ \bibinfo {author} {\bibfnamefont {H.}~\bibnamefont
			{Euler}},\ }\href {https://doi.org/10.1007/BF01343663} {\bibfield  {journal}
		{\bibinfo  {journal} {Zeitschrift f\"ur Physik}\ }\textbf {\bibinfo {volume}
			{98}},\ \bibinfo {pages} {714} (\bibinfo {year} {1936})}\BibitemShut
	{NoStop}%
	\bibitem [{\citenamefont {Nashed}\ and\ \citenamefont
		{Nojiri}(2021)}]{NashedNojiri2021}%
	\BibitemOpen
	\bibfield  {author} {\bibinfo {author} {\bibfnamefont {G.~G.~L.}\
			\bibnamefont {Nashed}}\ and\ \bibinfo {author} {\bibfnamefont
			{S.}~\bibnamefont {Nojiri}},\ }\href
	{https://doi.org/10.1103/PhysRevD.104.044043} {\bibfield  {journal} {\bibinfo
			{journal} {Phys. Rev. D}\ }\textbf {\bibinfo {volume} {104}},\ \bibinfo
		{pages} {044043} (\bibinfo {year} {2021})}\BibitemShut {NoStop}%
	\bibitem [{\citenamefont {Guerrero}\ and\ \citenamefont
		{Rubiera-Garcia}(2020)}]{GuerreroRubiera2020}%
	\BibitemOpen
	\bibfield  {author} {\bibinfo {author} {\bibfnamefont {M.}~\bibnamefont
			{Guerrero}}\ and\ \bibinfo {author} {\bibfnamefont {D.}~\bibnamefont
			{Rubiera-Garcia}},\ }\href {https://doi.org/10.1103/PhysRevD.102.024005}
	{\bibfield  {journal} {\bibinfo  {journal} {Phys. Rev. D}\ }\textbf {\bibinfo
			{volume} {102}},\ \bibinfo {pages} {024005} (\bibinfo {year}
		{2020})}\BibitemShut {NoStop}%
	\bibitem [{\citenamefont {Hamil}\ and\ \citenamefont
		{L\"utf\"uo\u{g}lu}(2024)}]{HamilLutfuoglu2024}%
	\BibitemOpen
	\bibfield  {author} {\bibinfo {author} {\bibfnamefont {B.}~\bibnamefont
			{Hamil}}\ and\ \bibinfo {author} {\bibfnamefont {B.~C.}\ \bibnamefont
			{L\"utf\"uo\u{g}lu}},\ }\href {https://doi.org/10.1002/prop.202400105}
	{\bibfield  {journal} {\bibinfo  {journal} {Fortsc. Phys.}\ }\textbf
		{\bibinfo {volume} {73}},\ \bibinfo {pages} {2400105} (\bibinfo {year}
		{2024})}\BibitemShut {NoStop}%
	\bibitem [{\citenamefont {Al-Badawi}\ and\ \citenamefont
		{Ahmed}(2025)}]{AlBadawiAhmed2025}%
	\BibitemOpen
	\bibfield  {author} {\bibinfo {author} {\bibfnamefont {A.}~\bibnamefont
			{Al-Badawi}}\ and\ \bibinfo {author} {\bibfnamefont {F.}~\bibnamefont
			{Ahmed}},\ }\href {https://doi.org/10.1016/j.cjph.2025.01.021} {\bibfield
		{journal} {\bibinfo  {journal} {Chin. J. Phys.}\ }\textbf {\bibinfo {volume}
			{94}},\ \bibinfo {pages} {185} (\bibinfo {year} {2025})}\BibitemShut
	{NoStop}%
	\bibitem [{\citenamefont {Ahmed}\ \emph
		{et~al.}(2025{\natexlab{a}})\citenamefont {Ahmed}, \citenamefont
		{Al-Badawi},\ and\ \citenamefont {Sakall{\i}}}]{AhmedBlackStrings2025}%
	\BibitemOpen
	\bibfield  {author} {\bibinfo {author} {\bibfnamefont {F.}~\bibnamefont
			{Ahmed}}, \bibinfo {author} {\bibfnamefont {A.}~\bibnamefont {Al-Badawi}},\
		and\ \bibinfo {author} {\bibfnamefont {{\.I}.}~\bibnamefont {Sakall{\i}}},\
	}\href {https://doi.org/10.1140/epjc/s10052-025-14266-y} {\bibfield
		{journal} {\bibinfo  {journal} {Eur. Phys. J C}\ }\textbf {\bibinfo {volume}
			{85}},\ \bibinfo {pages} {554} (\bibinfo {year}
		{2025}{\natexlab{a}})}\BibitemShut {NoStop}%
	\bibitem [{\citenamefont {Ahmed}\ \emph
		{et~al.}(2025{\natexlab{b}})\citenamefont {Ahmed}, \citenamefont
		{Al-Badawi},\ and\ \citenamefont {Sakall{\i}}}]{AhmedPDU2025}%
	\BibitemOpen
	\bibfield  {author} {\bibinfo {author} {\bibfnamefont {F.}~\bibnamefont
			{Ahmed}}, \bibinfo {author} {\bibfnamefont {A.}~\bibnamefont {Al-Badawi}},\
		and\ \bibinfo {author} {\bibfnamefont {{\.I}.}~\bibnamefont {Sakall{\i}}},\
	}\href {https://doi.org/10.1016/j.dark.2025.101988} {\bibfield  {journal}
		{\bibinfo  {journal} {Phys. Dark Univ.}\ }\textbf {\bibinfo {volume} {49}},\
		\bibinfo {pages} {101988} (\bibinfo {year} {2025}{\natexlab{b}})}\BibitemShut
	{NoStop}%
	\bibitem [{\citenamefont {Ahmed}\ \emph
		{et~al.}(2025{\natexlab{c}})\citenamefont {Ahmed}, \citenamefont {Al-Badawi},
		\citenamefont {Sakall{\i}},\ and\ \citenamefont {Bouzenada}}]{AhmedBTZ2025}%
	\BibitemOpen
	\bibfield  {author} {\bibinfo {author} {\bibfnamefont {F.}~\bibnamefont
			{Ahmed}}, \bibinfo {author} {\bibfnamefont {A.}~\bibnamefont {Al-Badawi}},
		\bibinfo {author} {\bibfnamefont {{\.I}.}~\bibnamefont {Sakall{\i}}},\ and\
		\bibinfo {author} {\bibfnamefont {A.}~\bibnamefont {Bouzenada}},\ }\href
	{https://doi.org/10.1016/j.nuclphysb.2025.116806} {\bibfield  {journal}
		{\bibinfo  {journal} {Nucl. Phys. B}\ }\textbf {\bibinfo {volume} {1011}},\
		\bibinfo {pages} {116806} (\bibinfo {year} {2025}{\natexlab{c}})}\BibitemShut
	{NoStop}%
	\bibitem [{\citenamefont {Breton}\ and\ \citenamefont
		{L{\'o}pez}(2021)}]{BretonLopez2021}%
	\BibitemOpen
	\bibfield  {author} {\bibinfo {author} {\bibfnamefont {N.}~\bibnamefont
			{Breton}}\ and\ \bibinfo {author} {\bibfnamefont {L.~A.}\ \bibnamefont
			{L{\'o}pez}},\ }\href {https://doi.org/10.1103/PhysRevD.104.024064}
	{\bibfield  {journal} {\bibinfo  {journal} {Phys. Rev. D}\ }\textbf {\bibinfo
			{volume} {104}},\ \bibinfo {pages} {024064} (\bibinfo {year}
		{2021})}\BibitemShut {NoStop}%
	\bibitem [{\citenamefont {Magos}\ and\ \citenamefont
		{Breton}(2020)}]{MagosBreton2020EHAdS}%
	\BibitemOpen
	\bibfield  {author} {\bibinfo {author} {\bibfnamefont {D.}~\bibnamefont
			{Magos}}\ and\ \bibinfo {author} {\bibfnamefont {N.}~\bibnamefont {Breton}},\
	}\href {https://doi.org/10.1103/PhysRevD.102.084011} {\bibfield  {journal}
		{\bibinfo  {journal} {Phys. Rev. D}\ }\textbf {\bibinfo {volume} {102}},\
		\bibinfo {pages} {084011} (\bibinfo {year} {2020})}\BibitemShut {NoStop}%
	\bibitem [{\citenamefont {Salazar~I.}\ \emph {et~al.}(1987)\citenamefont
		{Salazar~I.}, \citenamefont {Garc\'ia~D.},\ and\ \citenamefont
		{Pleba\'nski}}]{SalazarEtAl1987}%
	\BibitemOpen
	\bibfield  {author} {\bibinfo {author} {\bibfnamefont {H.}~\bibnamefont
			{Salazar~I.}}, \bibinfo {author} {\bibfnamefont {A.}~\bibnamefont
			{Garc\'ia~D.}},\ and\ \bibinfo {author} {\bibfnamefont {J.~F.}\ \bibnamefont
			{Pleba\'nski}},\ }\href {https://doi.org/10.1063/1.527430} {\bibfield
		{journal} {\bibinfo  {journal} {J Math. Phys}\ }\textbf {\bibinfo {volume}
			{28}},\ \bibinfo {pages} {2171} (\bibinfo {year} {1987})}\BibitemShut
	{NoStop}%
	\bibitem [{\citenamefont {Kiselev}(2003)}]{Kiselev2003}%
	\BibitemOpen
	\bibfield  {author} {\bibinfo {author} {\bibfnamefont {V.~V.}\ \bibnamefont
			{Kiselev}},\ }\href {https://doi.org/10.1088/0264-9381/20/6/310} {\bibfield
		{journal} {\bibinfo  {journal} {Class. Quantum Grav.}\ }\textbf {\bibinfo
			{volume} {20}},\ \bibinfo {pages} {1187} (\bibinfo {year}
		{2003})}\BibitemShut {NoStop}%
	\bibitem [{\citenamefont {Li}\ and\ \citenamefont {Yang}(2012)}]{LiYang2012}%
	\BibitemOpen
	\bibfield  {author} {\bibinfo {author} {\bibfnamefont {M.-H.}\ \bibnamefont
			{Li}}\ and\ \bibinfo {author} {\bibfnamefont {K.-C.}\ \bibnamefont {Yang}},\
	}\href {https://doi.org/10.1103/PhysRevD.86.123015} {\bibfield  {journal}
		{\bibinfo  {journal} {Phys. Rev. D}\ }\textbf {\bibinfo {volume} {86}},\
		\bibinfo {pages} {123015} (\bibinfo {year} {2012})}\BibitemShut {NoStop}%
	\bibitem [{\citenamefont {Letelier}(1979)}]{Letelier1979}%
	\BibitemOpen
	\bibfield  {author} {\bibinfo {author} {\bibfnamefont {P.~S.}\ \bibnamefont
			{Letelier}},\ }\href {https://doi.org/10.1103/PhysRevD.20.1294} {\bibfield
		{journal} {\bibinfo  {journal} {Phys. Rev. D}\ }\textbf {\bibinfo {volume}
			{20}},\ \bibinfo {pages} {1294} (\bibinfo {year} {1979})}\BibitemShut
	{NoStop}%
	\bibitem [{\citenamefont {Bezerra}\ \emph {et~al.}(2019)\citenamefont
		{Bezerra}, \citenamefont {Lobo}, \citenamefont {Morais~Gra\c{c}a},\ and\
		\citenamefont {Santos}}]{BezerraEtAl2019}%
	\BibitemOpen
	\bibfield  {author} {\bibinfo {author} {\bibfnamefont {V.~B.}\ \bibnamefont
			{Bezerra}}, \bibinfo {author} {\bibfnamefont {I.~P.}\ \bibnamefont {Lobo}},
		\bibinfo {author} {\bibfnamefont {J.~P.}\ \bibnamefont {Morais~Gra\c{c}a}},\
		and\ \bibinfo {author} {\bibfnamefont {L.~C.~N.}\ \bibnamefont {Santos}},\
	}\href {https://doi.org/10.1140/epjc/s10052-019-7482-0} {\bibfield  {journal}
		{\bibinfo  {journal} {Eur. Phys. J C}\ }\textbf {\bibinfo {volume} {79}},\
		\bibinfo {pages} {949} (\bibinfo {year} {2019})}\BibitemShut {NoStop}%
	\bibitem [{\citenamefont {Abbas}\ and\ \citenamefont
		{Ali}(2023)}]{AbbasAli2023}%
	\BibitemOpen
	\bibfield  {author} {\bibinfo {author} {\bibfnamefont {G.}~\bibnamefont
			{Abbas}}\ and\ \bibinfo {author} {\bibfnamefont {R.~H.}\ \bibnamefont
			{Ali}},\ }\href {https://doi.org/10.1140/epjc/s10052-023-11580-1} {\bibfield
		{journal} {\bibinfo  {journal} {Eur. Phys. J C}\ }\textbf {\bibinfo {volume}
			{83}},\ \bibinfo {pages} {407} (\bibinfo {year} {2023})}\BibitemShut
	{NoStop}%
	\bibitem [{\citenamefont {Sood}\ \emph {et~al.}(2024)\citenamefont {Sood},
		\citenamefont {Ali}, \citenamefont {Singh},\ and\ \citenamefont
		{Ghosh}}]{SoodEtAl2024}%
	\BibitemOpen
	\bibfield  {author} {\bibinfo {author} {\bibfnamefont {A.}~\bibnamefont
			{Sood}}, \bibinfo {author} {\bibfnamefont {M.~S.}\ \bibnamefont {Ali}},
		\bibinfo {author} {\bibfnamefont {J.~K.}\ \bibnamefont {Singh}},\ and\
		\bibinfo {author} {\bibfnamefont {S.~G.}\ \bibnamefont {Ghosh}},\ }\href
	{https://doi.org/10.1088/1674-1137/ad361f} {\bibfield  {journal} {\bibinfo
			{journal} {Chin. Phys. C}\ }\textbf {\bibinfo {volume} {48}},\ \bibinfo
		{pages} {065109} (\bibinfo {year} {2024})}\BibitemShut {NoStop}%
	\bibitem [{\citenamefont {Heydari-Fard}\ \emph {et~al.}(2023)\citenamefont
		{Heydari-Fard}, \citenamefont {Ghassemi~Honarvar},\ and\ \citenamefont
		{Heydari-Fard}}]{HeydariFardEtAl2023}%
	\BibitemOpen
	\bibfield  {author} {\bibinfo {author} {\bibfnamefont {M.}~\bibnamefont
			{Heydari-Fard}}, \bibinfo {author} {\bibfnamefont {S.}~\bibnamefont
			{Ghassemi~Honarvar}},\ and\ \bibinfo {author} {\bibfnamefont
			{M.}~\bibnamefont {Heydari-Fard}},\ }\href
	{https://doi.org/10.1093/mnras/stad558} {\bibfield  {journal} {\bibinfo
			{journal} {MNRAS}\ }\textbf {\bibinfo {volume} {521}},\ \bibinfo {pages}
		{708} (\bibinfo {year} {2023})}\BibitemShut {NoStop}%
	\bibitem [{\citenamefont {Hou}\ \emph {et~al.}(2018)\citenamefont {Hou},
		\citenamefont {Xu},\ and\ \citenamefont {Wang}}]{HouXuWang2018}%
	\BibitemOpen
	\bibfield  {author} {\bibinfo {author} {\bibfnamefont {X.}~\bibnamefont
			{Hou}}, \bibinfo {author} {\bibfnamefont {Z.}~\bibnamefont {Xu}},\ and\
		\bibinfo {author} {\bibfnamefont {J.}~\bibnamefont {Wang}},\ }\href
	{https://doi.org/10.1088/1475-7516/2018/12/040} {\bibfield  {journal}
		{\bibinfo  {journal} {JCAP}\ }\textbf {\bibinfo {volume} {2018}}\bibinfo
		{number} { (12)},\ \bibinfo {pages} {040}}\BibitemShut {NoStop}%
	\bibitem [{\citenamefont {Haroon}\ \emph {et~al.}(2019)\citenamefont {Haroon},
		\citenamefont {Jamil}, \citenamefont {Jusufi}, \citenamefont {Lin},\ and\
		\citenamefont {Mann}}]{HaroonEtAl2019}%
	\BibitemOpen
	\bibfield  {number} {  }\bibfield  {author} {\bibinfo {author} {\bibfnamefont
			{S.}~\bibnamefont {Haroon}}, \bibinfo {author} {\bibfnamefont
			{M.}~\bibnamefont {Jamil}}, \bibinfo {author} {\bibfnamefont
			{K.}~\bibnamefont {Jusufi}}, \bibinfo {author} {\bibfnamefont
			{K.}~\bibnamefont {Lin}},\ and\ \bibinfo {author} {\bibfnamefont {R.~B.}\
			\bibnamefont {Mann}},\ }\href {https://doi.org/10.1103/PhysRevD.99.044015}
	{\bibfield  {journal} {\bibinfo  {journal} {Phys. Rev. D}\ }\textbf {\bibinfo
			{volume} {99}},\ \bibinfo {pages} {044015} (\bibinfo {year}
		{2019})}\BibitemShut {NoStop}%
	\bibitem [{\citenamefont {Shahzad}\ \emph {et~al.}(2025)\citenamefont
		{Shahzad}, \citenamefont {Abbas}, \citenamefont {Zhu} \emph
		{et~al.}}]{ShahzadEtAl2025}%
	\BibitemOpen
	\bibfield  {author} {\bibinfo {author} {\bibfnamefont {M.~R.}\ \bibnamefont
			{Shahzad}}, \bibinfo {author} {\bibfnamefont {G.}~\bibnamefont {Abbas}},
		\bibinfo {author} {\bibfnamefont {T.}~\bibnamefont {Zhu}}, \emph {et~al.},\
	}\href {https://doi.org/10.1140/epjc/s10052-025-13876-w} {\bibfield
		{journal} {\bibinfo  {journal} {Eur. Phys. J. C}\ }\textbf {\bibinfo {volume}
			{85}},\ \bibinfo {pages} {164} (\bibinfo {year} {2025})}\BibitemShut
	{NoStop}%
	\bibitem [{\citenamefont {Rehman}\ \emph {et~al.}(2025)\citenamefont {Rehman},
		\citenamefont {Abbas}, \citenamefont {Zhu},\ and\ \citenamefont
		{Alkahtani}}]{Rehman2025AccretionEH}%
	\BibitemOpen
	\bibfield  {author} {\bibinfo {author} {\bibfnamefont {H.}~\bibnamefont
			{Rehman}}, \bibinfo {author} {\bibfnamefont {G.}~\bibnamefont {Abbas}},
		\bibinfo {author} {\bibfnamefont {T.}~\bibnamefont {Zhu}},\ and\ \bibinfo
		{author} {\bibfnamefont {B.~S.}\ \bibnamefont {Alkahtani}},\ }\href
	{https://doi.org/10.1016/j.dark.2025.101944} {\bibfield  {journal} {\bibinfo
			{journal} {Phys. Dark Univ.}\ }\textbf {\bibinfo {volume} {48}},\ \bibinfo
		{pages} {101944} (\bibinfo {year} {2025})}\BibitemShut {NoStop}%
	\bibitem [{\citenamefont {Rehman}\ \emph {et~al.}(2023)\citenamefont {Rehman},
		\citenamefont {Abbas}, \citenamefont {Zhu} \emph
		{et~al.}}]{Rehman2023AccretionEH}%
	\BibitemOpen
	\bibfield  {author} {\bibinfo {author} {\bibfnamefont {H.}~\bibnamefont
			{Rehman}}, \bibinfo {author} {\bibfnamefont {G.}~\bibnamefont {Abbas}},
		\bibinfo {author} {\bibfnamefont {T.}~\bibnamefont {Zhu}}, \emph {et~al.},\
	}\href {https://doi.org/10.1140/epjc/s10052-023-12033-5} {\bibfield
		{journal} {\bibinfo  {journal} {Eur. Phys. J C}\ }\textbf {\bibinfo {volume}
			{83}},\ \bibinfo {pages} {856} (\bibinfo {year} {2023})}\BibitemShut
	{NoStop}%
	\bibitem [{\citenamefont {Abbas}\ and\ \citenamefont
		{Rehman}(2023)}]{AbbasRehman2023AccretionEHAdS}%
	\BibitemOpen
	\bibfield  {author} {\bibinfo {author} {\bibfnamefont {G.}~\bibnamefont
			{Abbas}}\ and\ \bibinfo {author} {\bibfnamefont {H.}~\bibnamefont {Rehman}},\
	}\href {https://doi.org/10.1002/prop.202200205} {\bibfield  {journal}
		{\bibinfo  {journal} {Fortsc. Phys.}\ }\textbf {\bibinfo {volume} {71}},\
		\bibinfo {pages} {2200205} (\bibinfo {year} {2023})}\BibitemShut {NoStop}%
	\bibitem [{\citenamefont {Acan~Yildiz}\ \emph {et~al.}(2024)\citenamefont
		{Acan~Yildiz}, \citenamefont {Ditta}, \citenamefont {Ashraf}, \citenamefont
		{G{\"u}dekli}, \citenamefont {Alanazi},\ and\ \citenamefont
		{Reyimberganov}}]{AcanYildiz2024EH_PFDM}%
	\BibitemOpen
	\bibfield  {author} {\bibinfo {author} {\bibfnamefont {G.~D.}\ \bibnamefont
			{Acan~Yildiz}}, \bibinfo {author} {\bibfnamefont {A.}~\bibnamefont {Ditta}},
		\bibinfo {author} {\bibfnamefont {A.}~\bibnamefont {Ashraf}}, \bibinfo
		{author} {\bibfnamefont {E.}~\bibnamefont {G{\"u}dekli}}, \bibinfo {author}
		{\bibfnamefont {Y.~M.}\ \bibnamefont {Alanazi}},\ and\ \bibinfo {author}
		{\bibfnamefont {A.}~\bibnamefont {Reyimberganov}},\ }\href
	{https://doi.org/10.1016/j.dark.2024.101583} {\bibfield  {journal} {\bibinfo
			{journal} {Phys. Dark Univ.}\ }\textbf {\bibinfo {volume} {46}},\ \bibinfo
		{pages} {101583} (\bibinfo {year} {2024})}\BibitemShut {NoStop}%
	\bibitem [{\citenamefont {Chaudhary}\ \emph {et~al.}(2025)\citenamefont
		{Chaudhary}, \citenamefont {Sultan}, \citenamefont {Malik}, \citenamefont
		{Alanazi}, \citenamefont {Bin~Jumah}, \citenamefont {Ashraf},\ and\
		\citenamefont {Mubaraki}}]{Chaudhary2025EH_PFDM}%
	\BibitemOpen
	\bibfield  {author} {\bibinfo {author} {\bibfnamefont {S.}~\bibnamefont
			{Chaudhary}}, \bibinfo {author} {\bibfnamefont {M.~D.}\ \bibnamefont
			{Sultan}}, \bibinfo {author} {\bibfnamefont {A.}~\bibnamefont {Malik}},
		\bibinfo {author} {\bibfnamefont {Y.~M.}\ \bibnamefont {Alanazi}}, \bibinfo
		{author} {\bibfnamefont {A.}~\bibnamefont {Bin~Jumah}}, \bibinfo {author}
		{\bibfnamefont {A.}~\bibnamefont {Ashraf}},\ and\ \bibinfo {author}
		{\bibfnamefont {A.~M.}\ \bibnamefont {Mubaraki}},\ }\href
	{https://doi.org/10.1142/S0219887825501294} {\bibfield  {journal} {\bibinfo
			{journal} {Int. J Geom. Meth. Mod. Phys.}\ }\textbf {\bibinfo {volume}
			{22}},\ \bibinfo {pages} {2550129} (\bibinfo {year} {2025})}\BibitemShut
	{NoStop}%
	\bibitem [{\citenamefont {Sucu}\ and\ \citenamefont
		{Sakalli}(2026)}]{SucuSakalli2026EHAdS}%
	\BibitemOpen
	\bibfield  {author} {\bibinfo {author} {\bibfnamefont {E.}~\bibnamefont
			{Sucu}}\ and\ \bibinfo {author} {\bibfnamefont {I.}~\bibnamefont {Sakalli}},\
	}\href {https://doi.org/10.1016/j.dark.2026.102295} {\bibfield  {journal}
		{\bibinfo  {journal} {Physics of the Dark Universe}\ }\textbf {\bibinfo
			{volume} {52}},\ \bibinfo {pages} {102295} (\bibinfo {year}
		{2026})}\BibitemShut {NoStop}%
	\bibitem [{\citenamefont {{Event Horizon Telescope
				Collaboration}}(2019{\natexlab{a}})}]{Akiyama2019L1}%
	\BibitemOpen
	\bibfield  {author} {\bibinfo {author} {\bibnamefont {{Event Horizon
					Telescope Collaboration}}},\ }\href
	{https://doi.org/10.3847/2041-8213/ab0ec7} {\bibfield  {journal} {\bibinfo
			{journal} {Astrophys. J Lett.}\ }\textbf {\bibinfo {volume} {875}},\ \bibinfo
		{pages} {L1} (\bibinfo {year} {2019}{\natexlab{a}})}\BibitemShut {NoStop}%
	\bibitem [{\citenamefont {{Event Horizon Telescope
				Collaboration}}(2019{\natexlab{b}})}]{Akiyama2019L4}%
	\BibitemOpen
	\bibfield  {author} {\bibinfo {author} {\bibnamefont {{Event Horizon
					Telescope Collaboration}}},\ }\href
	{https://doi.org/10.3847/2041-8213/ab0e85} {\bibfield  {journal} {\bibinfo
			{journal} {Astrophys. J Lett.}\ }\textbf {\bibinfo {volume} {875}},\ \bibinfo
		{pages} {L4} (\bibinfo {year} {2019}{\natexlab{b}})}\BibitemShut {NoStop}%
	\bibitem [{\citenamefont {{Event Horizon Telescope
				Collaboration}}(2022)}]{Akiyama2022L12}%
	\BibitemOpen
	\bibfield  {author} {\bibinfo {author} {\bibnamefont {{Event Horizon
					Telescope Collaboration}}},\ }\href
	{https://doi.org/10.3847/2041-8213/ac6674} {\bibfield  {journal} {\bibinfo
			{journal} {Astrophys. J Lett.}\ }\textbf {\bibinfo {volume} {930}},\ \bibinfo
		{pages} {L12} (\bibinfo {year} {2022})}\BibitemShut {NoStop}%
	\bibitem [{\citenamefont {Perlick}\ and\ \citenamefont
		{Tsupko}(2022)}]{PerlickTsupko2022}%
	\BibitemOpen
	\bibfield  {author} {\bibinfo {author} {\bibfnamefont {V.}~\bibnamefont
			{Perlick}}\ and\ \bibinfo {author} {\bibfnamefont {O.~Y.}\ \bibnamefont
			{Tsupko}},\ }\href {https://doi.org/10.1016/j.physrep.2021.10.004} {\bibfield
		{journal} {\bibinfo  {journal} {Phys. Rep.}\ }\textbf {\bibinfo {volume}
			{947}},\ \bibinfo {pages} {1} (\bibinfo {year} {2022})}\BibitemShut {NoStop}%
	\bibitem [{\citenamefont {Got\={o}}(1971)}]{Goto1971}%
	\BibitemOpen
	\bibfield  {author} {\bibinfo {author} {\bibfnamefont {T.}~\bibnamefont
			{Got\={o}}},\ }\href {https://doi.org/10.1143/PTP.46.1560} {\bibfield
		{journal} {\bibinfo  {journal} {Prog. Theor. Phys.}\ }\textbf {\bibinfo
			{volume} {46}},\ \bibinfo {pages} {1560} (\bibinfo {year}
		{1971})}\BibitemShut {NoStop}%
	\bibitem [{\citenamefont {Ruffini}\ \emph {et~al.}(2013)\citenamefont
		{Ruffini}, \citenamefont {Wu},\ and\ \citenamefont {Xue}}]{Ruffini2013EEH}%
	\BibitemOpen
	\bibfield  {author} {\bibinfo {author} {\bibfnamefont {R.}~\bibnamefont
			{Ruffini}}, \bibinfo {author} {\bibfnamefont {Y.-B.}\ \bibnamefont {Wu}},\
		and\ \bibinfo {author} {\bibfnamefont {S.-S.}\ \bibnamefont {Xue}},\ }\href
	{https://doi.org/10.1103/PhysRevD.88.085004} {\bibfield  {journal} {\bibinfo
			{journal} {Phys. Rev. D}\ }\textbf {\bibinfo {volume} {88}},\ \bibinfo
		{pages} {085004} (\bibinfo {year} {2013})}\BibitemShut {NoStop}%
	\bibitem [{\citenamefont {Visser}\ \emph {et~al.}(2017)\citenamefont {Visser},
		\citenamefont {Gray}, \citenamefont {Schuster},\ and\ \citenamefont
		{Van-Brunt}}]{Visser2017Sparsity}%
	\BibitemOpen
	\bibfield  {author} {\bibinfo {author} {\bibfnamefont {M.}~\bibnamefont
			{Visser}}, \bibinfo {author} {\bibfnamefont {F.}~\bibnamefont {Gray}},
		\bibinfo {author} {\bibfnamefont {S.}~\bibnamefont {Schuster}},\ and\
		\bibinfo {author} {\bibfnamefont {A.}~\bibnamefont {Van-Brunt}},\ }in\ \href
	{https://doi.org/10.1142/9789813226609_0175} {\emph {\bibinfo {booktitle}
			{The Fourteenth Marcel Grossmann Meeting}}}\ (\bibinfo {year} {2017})\ pp.\
	\bibinfo {pages} {1724--1729}\BibitemShut {NoStop}%
	\bibitem [{\citenamefont {Ahmed}\ \emph
		{et~al.}(2026{\natexlab{a}})\citenamefont {Ahmed}, \citenamefont
		{Al-Badawi},\ and\ \citenamefont {Silva}}]{Ahmed2026ModMaxAdS}%
	\BibitemOpen
	\bibfield  {author} {\bibinfo {author} {\bibfnamefont {F.}~\bibnamefont
			{Ahmed}}, \bibinfo {author} {\bibfnamefont {A.}~\bibnamefont {Al-Badawi}},\
		and\ \bibinfo {author} {\bibfnamefont {E.~O.}\ \bibnamefont {Silva}},\ }\href
	{https://doi.org/10.1016/j.physletb.2026.140448} {\bibfield  {journal}
		{\bibinfo  {journal} {Phys. Lett. B}\ }\textbf {\bibinfo {volume} {876}},\
		\bibinfo {pages} {140448} (\bibinfo {year} {2026}{\natexlab{a}})}\BibitemShut
	{NoStop}%
	\bibitem [{\citenamefont {Ahmed}\ \emph
		{et~al.}(2026{\natexlab{b}})\citenamefont {Ahmed}, \citenamefont
		{Al-Badawi},\ and\ \citenamefont {Sakalli}}]{Ahmed2026SkyrmionBH}%
	\BibitemOpen
	\bibfield  {author} {\bibinfo {author} {\bibfnamefont {F.}~\bibnamefont
			{Ahmed}}, \bibinfo {author} {\bibfnamefont {A.}~\bibnamefont {Al-Badawi}},\
		and\ \bibinfo {author} {\bibfnamefont {I.}~\bibnamefont {Sakalli}},\ }\href
	{https://doi.org/10.48550/arXiv.2604.06259} {} (\bibinfo {year}
	{2026}{\natexlab{b}}),\ \Eprint {https://arxiv.org/abs/2604.06259}
	{arXiv:2604.06259 [gr-qc]} \BibitemShut {NoStop}%
	\bibitem [{\citenamefont {Ahmed}\ \emph
		{et~al.}(2026{\natexlab{c}})\citenamefont {Ahmed}, \citenamefont {Fathi},\
		and\ \citenamefont {Silva}}]{Ahmed2026KR_PFDM}%
	\BibitemOpen
	\bibfield  {author} {\bibinfo {author} {\bibfnamefont {F.}~\bibnamefont
			{Ahmed}}, \bibinfo {author} {\bibfnamefont {M.}~\bibnamefont {Fathi}},\ and\
		\bibinfo {author} {\bibfnamefont {E.~O.}\ \bibnamefont {Silva}},\ }\href
	{https://doi.org/10.48550/arXiv.2604.11357} {} (\bibinfo {year}
	{2026}{\natexlab{c}}),\ \Eprint {https://arxiv.org/abs/2604.11357}
	{arXiv:2604.11357 [gr-qc]} \BibitemShut {NoStop}%
	\bibitem [{\citenamefont {Ahmed}\ \emph
		{et~al.}(2026{\natexlab{d}})\citenamefont {Ahmed}, \citenamefont {Kala},\
		and\ \citenamefont {Silva}}]{Ahmed2026BumblebeeMonopole}%
	\BibitemOpen
	\bibfield  {author} {\bibinfo {author} {\bibfnamefont {F.}~\bibnamefont
			{Ahmed}}, \bibinfo {author} {\bibfnamefont {S.}~\bibnamefont {Kala}},\ and\
		\bibinfo {author} {\bibfnamefont {E.~O.}\ \bibnamefont {Silva}},\ }\href
	{https://doi.org/10.48550/arXiv.2604.00883} {} (\bibinfo {year}
	{2026}{\natexlab{d}}),\ \Eprint {https://arxiv.org/abs/2604.00883}
	{arXiv:2604.00883 [gr-qc]} \BibitemShut {NoStop}%
	\bibitem [{\citenamefont {Ahmed}\ and\ \citenamefont
		{Silva}(2026)}]{Ahmed2026BumblebeeTsallis}%
	\BibitemOpen
	\bibfield  {author} {\bibinfo {author} {\bibfnamefont {F.}~\bibnamefont
			{Ahmed}}\ and\ \bibinfo {author} {\bibfnamefont {E.~O.}\ \bibnamefont
			{Silva}},\ }\href {https://doi.org/10.48550/arXiv.2603.19034} {} (\bibinfo
	{year} {2026}),\ \Eprint {https://arxiv.org/abs/2603.19034} {arXiv:2603.19034
		[gr-qc]} \BibitemShut {NoStop}%
	\bibitem [{\citenamefont {Gray}\ \emph {et~al.}(2016)\citenamefont {Gray},
		\citenamefont {Schuster}, \citenamefont {Van-Brunt},\ and\ \citenamefont
		{Visser}}]{Gray2016}%
	\BibitemOpen
	\bibfield  {author} {\bibinfo {author} {\bibfnamefont {F.}~\bibnamefont
			{Gray}}, \bibinfo {author} {\bibfnamefont {S.}~\bibnamefont {Schuster}},
		\bibinfo {author} {\bibfnamefont {A.}~\bibnamefont {Van-Brunt}},\ and\
		\bibinfo {author} {\bibfnamefont {M.}~\bibnamefont {Visser}},\ }\href
	{https://doi.org/10.1088/0264-9381/33/11/115003} {\bibfield  {journal}
		{\bibinfo  {journal} {Class. Quantum Grav.}\ }\textbf {\bibinfo {volume}
			{33}},\ \bibinfo {pages} {115003} (\bibinfo {year} {2016})}\BibitemShut
	{NoStop}%
	\bibitem [{\citenamefont {Page}(1976)}]{Page1976}%
	\BibitemOpen
	\bibfield  {author} {\bibinfo {author} {\bibfnamefont {D.~N.}\ \bibnamefont
			{Page}},\ }\href {https://doi.org/10.1103/PhysRevD.13.198} {\bibfield
		{journal} {\bibinfo  {journal} {Phys. Rev. D}\ }\textbf {\bibinfo {volume}
			{13}},\ \bibinfo {pages} {198} (\bibinfo {year} {1976})}\BibitemShut
	{NoStop}%
	\bibitem [{\citenamefont {Ahmed}\ \emph
		{et~al.}(2025{\natexlab{d}})\citenamefont {Ahmed}, \citenamefont
		{Al-Badawi},\ and\ \citenamefont {Sakall{\i}}}]{AhmedHernquist2025}%
	\BibitemOpen
	\bibfield  {author} {\bibinfo {author} {\bibfnamefont {F.}~\bibnamefont
			{Ahmed}}, \bibinfo {author} {\bibfnamefont {A.}~\bibnamefont {Al-Badawi}},\
		and\ \bibinfo {author} {\bibfnamefont {{\.I}.}~\bibnamefont {Sakall{\i}}},\
	}\href {https://doi.org/10.1140/epjc/s10052-025-14723-8} {\bibfield
		{journal} {\bibinfo  {journal} {Eur. Phys. J C}\ }\textbf {\bibinfo {volume}
			{85}},\ \bibinfo {pages} {984} (\bibinfo {year}
		{2025}{\natexlab{d}})}\BibitemShut {NoStop}%
	\bibitem [{\citenamefont {Wei}\ and\ \citenamefont {Liu}(2013)}]{WeiLiu2013}%
	\BibitemOpen
	\bibfield  {author} {\bibinfo {author} {\bibfnamefont {S.-W.}\ \bibnamefont
			{Wei}}\ and\ \bibinfo {author} {\bibfnamefont {Y.-X.}\ \bibnamefont {Liu}},\
	}\href {https://doi.org/10.1088/1475-7516/2013/11/063} {\bibfield  {journal}
		{\bibinfo  {journal} {JCAP}\ }\textbf {\bibinfo {volume} {2013}}\bibinfo
		{number} { (11)},\ \bibinfo {pages} {063}}\BibitemShut {NoStop}%
	\bibitem [{\citenamefont {D{\'e}canini}\ \emph {et~al.}(2011)\citenamefont
		{D{\'e}canini}, \citenamefont {Esposito-Far{\`e}se},\ and\ \citenamefont
		{Folacci}}]{DecaniniEtAl2011}%
	\BibitemOpen
	\bibfield  {number} {  }\bibfield  {author} {\bibinfo {author} {\bibfnamefont
			{Y.}~\bibnamefont {D{\'e}canini}}, \bibinfo {author} {\bibfnamefont
			{G.}~\bibnamefont {Esposito-Far{\`e}se}},\ and\ \bibinfo {author}
		{\bibfnamefont {A.}~\bibnamefont {Folacci}},\ }\href
	{https://doi.org/10.1103/PhysRevD.83.044032} {\bibfield  {journal} {\bibinfo
			{journal} {Phys. Rev. D}\ }\textbf {\bibinfo {volume} {83}},\ \bibinfo
		{pages} {044032} (\bibinfo {year} {2011})}\BibitemShut {NoStop}%
	\bibitem [{\citenamefont {Mashhoon}(1973)}]{Mashhoon1973}%
	\BibitemOpen
	\bibfield  {author} {\bibinfo {author} {\bibfnamefont {B.}~\bibnamefont
			{Mashhoon}},\ }\href {https://doi.org/10.1103/PhysRevD.7.2807} {\bibfield
		{journal} {\bibinfo  {journal} {Phys. Rev. D}\ }\textbf {\bibinfo {volume}
			{7}},\ \bibinfo {pages} {2807} (\bibinfo {year} {1973})}\BibitemShut
	{NoStop}%
	\bibitem [{\citenamefont {Sanchez}(1978)}]{Sanchez1978}%
	\BibitemOpen
	\bibfield  {author} {\bibinfo {author} {\bibfnamefont {N.}~\bibnamefont
			{Sanchez}},\ }\href {https://doi.org/10.1103/PhysRevD.18.1030} {\bibfield
		{journal} {\bibinfo  {journal} {Phys. Rev. D}\ }\textbf {\bibinfo {volume}
			{18}},\ \bibinfo {pages} {1030} (\bibinfo {year} {1978})}\BibitemShut
	{NoStop}%
\end{thebibliography}

\end{document}